\documentclass[11pt,preprint,showpacs]{revtex4}
\usepackage{graphicx}
\usepackage{dcolumn}
\usepackage{bm}
\begin{document}

\title{Three-nucleon force in relativistic
  three-nucleon Faddeev calculations}

\author{H.~Wita{\l}a}
\affiliation{M. Smoluchowski Institute of Physics, Jagiellonian
University,  PL-30059 Krak\'ow, Poland}

\author{J.~Golak}
\affiliation{M. Smoluchowski Institute of Physics, Jagiellonian
University,  PL-30059 Krak\'ow, Poland}

\author{R.~Skibi\'nski}
\affiliation{M. Smoluchowski Institute of Physics, Jagiellonian
University,  PL-30059 Krak\'ow, Poland}

\author{W.\ Gl\"ockle}
\affiliation{Institut f\"ur theoretische Physik II,
Ruhr-Universit\"at Bochum, D-44780 Bochum, Germany}

\author{H.\ Kamada}
\affiliation{Department of Physics, Faculty of Engineering,
Kyushu Institute of Technology, Kitakyushu 804-8550, Japan}

\author{W. N.\ Polyzou}
\affiliation{Department of Physics and Astronomy,
The University of Iowa, Iowa City, IA 52242}

\date{\today}

\begin{abstract}

  We extend our formulation of relativistic three-nucleon Faddeev
  equations to include both pairwise interactions and a three-nucleon
  force.  Exact Poincar\'e invariance is realized by adding
  interactions to the mass Casimir operator (rest Hamiltonian) of the
  non-interacting system without changing the spin Casimir operator.
  This is achieved by using interactions defined by rotationally
  invariant kernels that are functions of internal momentum variables
  and single-particle spins that undergo identical Wigner rotations.
  To solve the resulting equations one needs matrix elements of the
  three-nucleon force with these properties in a momentum-space
  partial-wave basis.  We present two methods to calculate matrix
  elements of three-nucleon forces with these properties.  For a
  number of examples we show that at higher energies, where effects of
  relativity and of three-nucleon forces are non-negligible, a
  consistent treatment of both is required to properly analyze the
  data.

\end{abstract}


\maketitle
\setcounter{page}{1}

\section{Introduction}
\label{intro}

High precision nucleon-nucleon potentials such as AV18~\cite{AV18},
CDBonn~\cite{CDBOnucleon-nucleon}, Nijm I, II and 93~\cite{NIJMI}
provide a very good description of the nucleon-nucleon data set up to
about 350 MeV.  When these forces are used to predict binding energies
of three-nucleon systems they underestimate the experimental bindings
of $^3H$ and $^3He$ by about 0.5-1
MeV~\cite{Friar1993,Nogga1997}. This missing binding energy can be
restored by introducing a three-nucleon force into the nuclear
Hamiltonian~\cite{Nogga1997}.

Also the study of elastic nucleon-deuteron scattering and nucleon
induced deuteron breakup revealed a number of cases where the
nonrelativistic description using only pairwise forces is insufficient
to explain the data.  Generally, the studied discrepancies between a
theory using only nucleon-nucleon potentials and experiment become
larger with increasing energy of the three-nucleon system. Adding a
three-nucleon force to the pairwise interactions leads in some cases
to a better description of the data.  The elastic nucleon-deuteron
angular distribution in the region of its minimum and at backward
angles is the best studied example~\cite{wit98,sek02}.  The clear
discrepancy in these angular regions at energies up to
$\approx100$~MeV nucleon lab energy between a theory using only
nucleon-nucleon potentials and the cross section data can be removed
by adding a modern three-nucleon force to the nuclear
Hamiltonian. Such a three-nucleon force must be adjusted with each
nucleon-nucleon potential separately to the experimental binding of
$^3H$ and $^3He$~\cite{wit98,wit01,sek02}.  At energies higher than
$\approx 100$~MeV current three-nucleon forces only partially improve
the description of cross section data and the remaining discrepancies,
which increase with energy, indicate the possibility of relativistic
effects.  The need for a relativistic description of three-nucleon
scattering was also raised when precise measurements of the total
cross section for neutron-deuteron scattering~\cite{abf98} were
analyzed within the framework of nonrelativistic Faddeev
calculations~\cite{wit99}.  Nucleon-nucleon forces alone were
insufficient to describe the data above $\approx 100$~MeV.  The
effects due to relativistic kinematics considered in \cite{wit99} were
comparable at higher energies to the effects due to three-nucleon
forces.  These results showed the importance of a study taking
relativistic effects in the three nucleon continuum into account.

In \cite{witrel1,witrel2} the first results on relativistic effects in
the three-nucleon continuum have been presented.  The dynamics was
defined by a three-nucleon center of momentum Hamiltonian or mass
operator including only pairwise interactions.  The mass operator was
used to calculate three-nucleon scattering observables.  The input to
that approach is a ``Lorentz boosted'' nucleon-nucleon potential,
which generates the nucleon-nucleon $t$-matrix in a moving frame by
solving a standard Lippmann-Schwinger equation.  To get the
nucleon-nucleon potential in an arbitrary moving frame one needs the
interaction in the two-nucleon center of momentum system, which
appears in the relativistic nucleon-nucleon Schr\"odinger or
Lippmann-Schwinger equation.  The relativistic Schr\"odinger equation 
 in the two-nucleon center of momentum system 
 differs from the nonrelativistic Schr\"odinger equation just by the
relativistic form for the kinetic energy.  Current realistic
nucleon-nucleon potentials are defined and fit by comparing the
solution of the nonrelativistic Schr\"odinger equation to experimental
data.  Up to now nucleon-nucleon potentials refitted with the same
accuracy in the framework of the relativistic nucleon-nucleon
Schr\"odinger equation do not exist. Such refitting can be, however,
avoided by solving a quadratic integral equation whose solution is a
relativistic potential which is phase-equivalent to a given input
high-precision nonrelativistic nucleon-nucleon potential
\cite{kamada1}.  An alternative equivalent approach towards a relativistic 
nucleon-nucleon t-matrix in another frame is provided in \cite{CSP}.

In our previous studies with only nucleon-nucleon interactions we found
that when the non-relativistic form of the kinetic energy is replaced
by the relativistic one and a proper treatment of the relativistic dynamics is
included, the elastic scattering cross section is only slightly
influenced by relativity.  Only at backward angles and higher energies
are the elastic cross sections increased by relativity
\cite{witrel1}. It is exactly the region of angles and energies where
the effects of three-nucleon forces are also significant
\cite{wit01}. Also, for nucleon-deuteron breakup reactions regions of
phase space were found at higher energies of the incoming neutron
where relativity significantly changes the breakup cross sections
\cite{witrel3,skibbr}. For some spin observables large effects due to
relativity and three-nucleon forces have been reported in
nucleon-deuteron breakup for an incoming deuteron energy of $270$~MeV,
some of which seem to be supported by proton-deuteron data
\cite{brkim}.  These observations call for three-nucleon continuum
relativistic Faddeev calculations which include three-nucleon
forces. Only such consistent calculations should be used to analyze
the data in cases when both relativity and three-nucleon force effects
are large.

The paper is organized as follows.  
 Sec.~\ref{rel_dyn}  provides the conceptual basis
for the choice of the momentum-space representation and the definition
of spin in the relativistic context. 
 In Sec.~\ref{form} we summarize
the formalism underlying relativistic three-nucleon Faddeev
calculations with only nucleon-nucleon interactions, presented in
detail in \cite{witrel1,witrel2}.  In Sec.~\ref{3nfmatrixelement} we
focus on the three-nucleon Faddeev equation with an included
three-nucleon force and discuss two methods to compute matrix elements
of the three-nucleon force in the partial wave basis used in our
relativistic calculations.  In Sec.~\ref{results} we apply our
formulation to elastic nucleon-deuteron scattering and breakup and
show and discuss the results.  Sec.~\ref{summary} contains our
conclusions and summary. 
 Appendixes \ref{app_a} and \ref{app_b} formulate three-nucleon forces in the
 momentum  space representation adapted to Poincar\'e 
 invariance. 

\section{Relativistic dynamics}
\label{rel_dyn}

Relativistic invariance of a quantum theory means that the Poincar\'e
group (inhomogeneous Lorentz group) is a symmetry group of
the theory.  This requires the existence of a unitary representation
of the Poincar\'e group \cite{Wigner:1939cj}.  The Poincar\'e group
has ten generators, six Lorentz generators $J^{\mu \nu}$, and four
spacetime translation generators, $P^{\mu}$.  The dynamics of the
system is given by the Hamiltonian, $H=P^0$.  The Lie algebra has two
polynomial invariants,

\begin{equation} 
M^2 = - P^{\mu}P_{\mu}
\qquad 
W^2 = W^{\mu}W_{\mu} 
\label{bb.1}
\end{equation} 
where $W^{\mu}$ is the Pauli-Lubanski vector \cite{paulilubanski}  
\begin{equation}
W^{\mu}:=   
- {1 \over 2} \epsilon^{\mu \alpha \beta \gamma}P_{\alpha}J_{\beta \gamma}.
\label{bb.2}
\end{equation}
It satisfies
\begin{equation}
[P^{\mu},W^{\nu}]=0, \qquad P^{\mu}W_{\mu}=0, \qquad [W^{\mu},W^{\nu}]=
i \epsilon^{\mu\nu\alpha\beta}P_{\alpha} W_{\beta} .
\label{bb.3}
\end{equation}
Equation (\ref{bb.1}) implies that the Hamiltonian can be expressed in
terms of the mass operator, $H=\sqrt{M^2 + {P}^2}$, where $P^2:=
\mathbf{P}^2$.  Thus, given a representation for $\mathbf{P}$, the
dynamics is defined by the mass operator $M$, which plays the same
role in Poincar\'e invariant quantum mechanics as the center of mass
Hamiltonian $h= H-{P}^2/2M$ does in Galilean invariant quantum
mechanics.

In the absence of interactions the mass operator $M$ becomes the
invariant mass operator $M_0$ of three non-interacting relativistic 
particles.  The full interaction is defined by 
\begin{equation}
V := M - M_0 .
\label{bb.4}
\end{equation} 

For a system of three particles interacting with short-range
interactions, two-body interactions are defined by
\begin{equation} 
V_{(ij)(k)} := M_{(ij)(k)} - M_0 
\label{bb.5}
\end{equation} 
where $M_{(ij)(k)}$ is obtained from $M$ by turning off all interactions
in $M$ that involve particle $k$.  The difference 
\begin{equation} 
V_{4} := V-V_{(12)(3)}- V_{(23)(1)}-V_{(31)(2)} 
\label{bb.6}
\end{equation} 
defines a three-body interaction.  With these definitions the 
mass operator has the form 
\begin{equation}
M = M_0 + V_{(12)(3)}+ V_{(23)(1)}+V_{(31)(2)} + V_4 .
\label{bb.7}
\end{equation} 
This has the same form as the non-relativistic three-body center of mass
Hamiltonian with two and three-body forces, except the
non-relativistic kinetic energy is replaced by the relativistic
invariant mass of the non-interacting system.  As in the
non-relativistic case, bound and scattering eigenstates of this mass
operator can be computed using the Faddeev equations with two and
three-body interactions.  For identical nucleons the coupled
relativistic Faddeev equations can be replaced by a single equation.
Details are discussed in the next section.

In addition to the constraints imposed by discrete symmetries,
translational invariance, and particle exchange symmetry, there are
non-trivial constraints on the interactions due to both the Poincar\'e
symmetry and cluster properties.  The constraints on the interaction
due to Poincar\'e invariance come from the commutator 
\begin{equation}
[P^j ,J^{0k}] = i\delta_{jk}H, 
\label{bb.8}
\end{equation}
which means that interactions appearing in $H$ must be generated by
the operators in the commutator.  One way to satisfy the constraints
due to Poincar\'e invariance was suggested by Bakamjian and Thomas
\cite{Bakamjian:1953kh}. Their construction adds interactions to the 
mass Casimir operator that commute with the spin Casimir operator
\begin{equation}
\mathbf{j}^2 := W^2 /M^2 .
\end{equation}
The required interactions commute with and are independent of the
total momentum and commute with the non-interacting three-body
canonical spin operator.

Spin is associated with rotational degrees of freedom that appear in
the rest frame.  Because the Lorentz boost generators, $J^{0i}$, do
not form a closed sub-algebra, a sequence of Lorentz boosts that map
the rest frame to the rest frame can generate a rotation.  Thus in
order to obtain a well-defined relativistic spin it is necessary to
define a standard procedure for measuring the spin.  This normally
requires the specification of a special frame where spins can be
compared (usually the rest frame) and a standard set of Lorentz
transformations $B^{-1}(P)^{\mu}{}_{\nu}$, parameterized by momentum,
that transform arbitrary frames to the special frame.  The three-body
canonical spin is defined in terms of the Pauli-Lubanski vector by
\cite{kei91}
\begin{equation}
(0,\mathbf{j}_c)^{\mu} := {1 \over M}  B_c^{-1}(P)^{\mu}{}_{\nu} W^{\nu} 
\label{bb.9}
\end{equation}
where  $B_c^{-1}(P)^{\mu}{}_{\nu}$ is the rotationless Lorentz 
transformation-valued function of the four momentum $P$, 
\begin{equation}
B_c^{-1}(P)^{\mu}{}_{\nu} :=
\left (
\begin{array}{cc}
P^0/M  & -\mathbf{P}/M   \\
-\mathbf{P}/M & I + {\mathbf{P} \otimes \mathbf{P} \over M (P^0+M)}
\end{array} 
\right ).
\label{bb.10}
\end{equation}
This Lorentz transformation (\ref{bb.10}) satisfies
\begin{equation}
B_c^{-1}(P)^{\mu}{}_{\nu} P^{\nu} = (M,\mathbf{0})^{\mu}.
\label{bb.11}
\end{equation}
Equations (\ref{bb.3}) and (\ref{bb.11}) can be used to show that the
components of $\mathbf{j}_c$ satisfy $SU(2)$ commutation relations.
The spin $(0,\mathbf{j}_{c})^{\mu}$ is not a four vector because
$B_c^{-1}(P)^{\mu}{}_{\nu}$ is a matrix of operators, rather than a
constant Lorentz transformation.  Under Lorentz transformation the
canonical spin Wigner rotates
\begin{equation}
(0,\mathbf{j}'_{c})^{\mu} := R_{wc}(\Lambda ,P)^{\mu}{}_{\nu} 
(0,\mathbf{j}_{c})^{\nu} 
\label{bb.12}
\end{equation}
where $R_{wc}(\Lambda ,P)^{\mu}{}_{\nu} := (B_c^{-1}(\Lambda P)
\Lambda B_c(P))^{\mu}{}_{\nu}$.  The spin Casimir operator
$\mathbf{j}^2 = \mathbf{j}_c\cdot \mathbf{j}_c$ is independent of the
choice of boost (\ref{bb.10}) used to define the spin.  The
non-interacting (kinematic) canonical spin, $\mathbf{j}_{c0}$, is
obtained from (\ref{bb.9}) by replacing $M \to M_0$, $W^{\mu} \to
W^{\mu}_0$ in (\ref{bb.9}) and $M \to M_0$ in (\ref{bb.10}).  Thus,
Poincar\'e invariance can be satisfied provided the interactions
$V_{(ij)(k)}$ and $V_4$ commute with $\mathbf{j}_{c0}$.

The other non-trivial constraint on the interactions is imposed by
cluster properties.  The problem arises due to the non-linear relation
between the two-body interaction $v_{ij}$ in the two-body problem and
the corresponding two-body interaction, $V_{(ij)(k)}$, in the
three-body problem.  Cluster properties relate $V_{(ij)(k)}$ to the
Poincar\'e generators for the interacting $ij$ pair and spectator $k$.
Unfortunately each $2+1$ mass operator constructed by requiring
cluster properties commutes with a different spin Casimir operator,
which means that linear combinations of these interactions will break
Poincar\'e invariance.  Coester \cite{relform1} observed that these
interactions could be replaced by phase-equivalent interactions that
commute with $\mathbf{j}_{c0}$.  These interactions are designed to
satisfy cluster properties in the three-body rest frame.  Using the
Bakamjian-Thomas construction linear combinations of the phase
equivalent $V_{(ij)(k)}$'s can be added in a manner that preserves the
overall Poincar\'e invariance.  While these interactions do not lead
to generators that satisfy cluster properties, cluster properties in
the three-body rest frame and Poincar\'e invariance of the $S$ matrix
ensures that the three-body $S$-matrix retains cluster properties in
all frames.

To construct two-body interactions, the two-body interactions in the
two-body problem that commute with the two-body canonical spin are
replaced by phase equivalent two-body interactions in the three-body
problem that commute with the three-body canonical spin.  The phase
equivalent interactions are identified in the rest frame of the
three-body system.  They are determined in all other frames by the
requirement that the three-body spin remains kinematic (in the
Bakamjian-Thomas construction this choice fixes the representation of
the boost generators).

To construct interactions that commute with the three-body kinematic
canonical spin it is useful to introduce momenta and spin variables
that have the same Wigner rotation properties as the three-body
kinematic canonical spin.  This is because the kinematic canonical
spin can be constructed out of these degrees of freedom using
conventional methods for adding angular momenta.

The desired momentum operators are the relativistic analog of Jacobi
momenta.  In the non-relativistic case Jacobi momenta can be defined
using Galilean boosts to the two and three-body rest frames.  In the
relativistic case the Galilean boosts are replaced by the rotationless
boost (\ref{bb.10}) and the relevant Jacobi momenta are \cite{relform1}
\begin{equation}
q_i^{\mu} = B^{-1}_c (P)^{\mu}{}_{\nu} p_i^{\nu} \qquad P^{\mu} = p_1^{\mu} + 
p_2^{\mu} +p_3^{\mu}
\qquad q_i^{\mu} = (\sqrt{{q}_i^2+m^2},\mathbf{q}_i)
\label{bb.13}
\end{equation}
\begin{equation}
k_{ij}^{\mu} = B^{-1}_c (q_{ij})^{\mu}{}_{\nu} q_i^{\nu}  \qquad q_{ij}^{\mu}:= 
q_i^{\mu} +q_j^{\mu} .
\label{bb.14}
\end{equation}  

In terms of these variables 
\begin{equation}
M_0= \sum_{i=1}^3 \sqrt{m^2 + {q}_i^2} = 
\sqrt{m_{ij0}^2 + {q}_k^2} + \sqrt{m^2 + {q}_k^2}
\label{bb.15}
\end{equation}
where the two-body invariant mass is 
\begin{equation}
m_{ij0} = \sqrt{- (q_i + q_j)^{\mu}
(q_i + q_j)_{\mu}} = \sqrt{m_i^2 + {k}_{ij}^2}
+ \sqrt{m_j^2 + {k}_{ji}^2}.
\label{bb.16}
\end{equation}
The vector variables satisfy 
\begin{equation}
\sum_{i=1}^3 \mathbf{q}_i = \mathbf{0} \qquad 
\mathbf{k}_{ij} + \mathbf{k}_{ji}= \mathbf{0} .
\label{bb.17}
\end{equation}
The relevant property of these momentum vectors is that they
experience the same Wigner rotations as the three-body kinematic 
canonical spin
(\ref{bb.12}), 
\begin{equation}
q_i^{\mu} \to q_i^{\mu\prime} =  (B^{-1}_c (\Lambda P) \Lambda B_c (P) 
B^{-1}_c (P))^{\mu}{}_{\nu} p_i^{\nu} = R_{wc}(\Lambda, P)^{\mu}{}_{\nu} 
q_i^{\nu} ~. 
\label{bb.18}
\end{equation}
Similarly, 
\[
k_{ij}^{\mu} \to 
k_{ij}^{\mu\prime} = B^{-1}_c (q'_{ij})^{\mu}{}_{\nu} q_i^{\nu\prime}  =
(B^{-1}_c (R_{wc}(\Lambda, P) q_{ij}) R_{wc}(\Lambda, P))^{\mu}{}_{\nu} 
q_i^{\nu} =
\]
\begin{equation}
(B^{-1}_c (R_{wc}(\Lambda, P) q_{ij}) R_{wc}(\Lambda, P)
B_c (q_{ij}))^{\mu}{}_{\nu}  k_{ij}^{\nu} = 
R_{wc}(\Lambda, P)^{\mu}{}_{\nu} k_{ij}^{\nu}
\label{bb.20}
\end{equation}   
where the last line follows from the property of the rotationless
boosts (\ref{bb.10}) that the Wigner rotation of a rotation is the rotation
\cite{kei91}
\begin{equation}
R_{wc}(R,P)^{\mu}{}_{\nu} = R^{\mu}{}_{\nu}
\label{bb.21}
\end{equation}
for any $\mathbf{P}$.  Thus the $\mathbf{q}_i$ and $\mathbf{k}_{ij}$ all
undergo the same Wigner rotations as the three-body kinematic 
canonical spin.

Next we introduce single-particle spins with the same property.
Single-particle canonical spins can be constructed from 
single-particle Poincar\'e generators using
\begin{equation}
(0,\mathbf{j}_{ci})^{\mu} := {1 \over m}  B_c^{-1}(p_i)^{\mu}{}_{\nu} W_i^{\nu} 
\label{bb.22}
\end{equation}
where the operators on the right side of (\ref{bb.22}) are constructed by
replacing all of the three-body generators in (\ref{bb.2}), (\ref{bb.9}) and
(\ref{bb.10}) by the corresponding one-body generators.

Under kinematic Lorentz transformations the single-particle canonical spins
experience Wigner rotations, $R_{wc}(\Lambda ,p_i)^{\mu}{}_{\nu}$, that  
depend on the single-particle momenta.  These rotations differ from the 
Wigner rotations experienced by $\mathbf{q}_i$, $\mathbf{k}_{ij}$
and $\mathbf{j}_{c0}$.  This can be changed by introducing new
single-particle spin operators that replace the rotationless boost 
in (\ref{bb.22}) 
by a two step boost,
\begin{equation}
B^{-1}_c(p_i)^{\mu}{}_{\nu} \to 
(B^{-1}_c(q_i) B^{-1}_c(P))^{\mu}{}_{\nu} .
\label{bb.23}
\end{equation}
These two boosts agree when $\mathbf{P}=0$.
Note that both of these boosts transform $p_i^{\mu} \to
(m,0,0,0)^{\mu}$, so they differ by momentum dependent rotations.  We
call these spins three-body constituent spins to distinguish them from
single-particle canonical spins.  The constituent spin operators are
defined by \cite{relform1}
\begin{equation}
(0,\mathbf{j}_{3csi})^{\mu} := {1 \over m} 
(B_c^{-1}(q_i) B_c^{-1}(P)) ^{\mu}{}_{\nu} W_i^{\nu} .
\label{bb.24}
\end{equation}
When $\mathbf{P}=0$, $B_c^{-1}(p_i) \to B_c^{-1}(q_i)$ which means
that single-particle canonical spins and three-body constituent spins
agree in the three-body rest frame.  For a three-body system the total
spin is identified with total angular momentum in the three-body rest
frame, which is the sum of the single-particle angular momenta. The
angular momentum of a single particle in the three-body rest frame is
the sum of the single-particle constituent spin and a contribution
from the single particle orbital angular momenta.

A calculation, using the property (\ref{bb.21}), shows that under Lorentz
transformations 
\begin{equation}
(0,\mathbf{j}'_{3csi})^{\mu} := R_{wc}(\Lambda,P)^{\mu}{}_{\nu}
(0,\mathbf{j}_{3csi})^{\nu} ~,
\label{bb.25}
\end{equation}
Wigner rotates with the same rotation as the vectors $\mathbf{q}_i$
and $\mathbf{k}_{ij}$ and the three-body kinematic canonical spin.
The three-body kinematic canonical spin is the sum of the orbital
angular momenta associated with $\mathbf{q}_k$ and $\mathbf{k}_{ij}$
and the single-particle three-body constituent spins.  The requirement
that an interaction commutes with the kinematic three-body canonical
spin is equivalent to the requirement that the interaction have a
rotationally invariant kernel when expressed in terms of these
variables.  Thus the required interactions in the Bakamjian-Thomas
construction are given by kernels of the form
\begin{equation}
\langle \mathbf{P}, \mathbf{q}_i, \mathbf{k}_{jk}, \mu_1, \mu_2, \mu_3 \vert V 
\vert \mathbf{P}', \mathbf{q}'_i, \mathbf{k}'_{jk}, \mu_1', \mu_2', \mu_3' 
\rangle 
= \delta(\mathbf{P}-\mathbf{P}')
\langle \mathbf{q}_i, \mathbf{k}_{jk}, \mu_1, \mu_2, \mu_3 \Vert V 
\Vert  \mathbf{q}'_i, \mathbf{k}'_{jk}, \mu_1', \mu_2', \mu_3' 
\rangle 
\label{bb.25a}
\end{equation}
where the reduced kernel is a rotationally-invariant function of 
$\mathbf{q}_i, \mathbf{k}_{jk}$ and the three-body 
constituent spins.

Two-body interactions in the two-body problem $v_{12}$ 
have a similar form
\begin{equation}
\langle \mathbf{P}_{12} ,\mathbf{k}_{12} , \mu_1, \mu_2 \vert 
v_{12} \vert \mathbf{P}_{12}' ,\mathbf{k}_{12}', \mu_1', \mu_2'  \rangle =
\delta (\mathbf{P}_{12} - \mathbf{P}_{12}' ) \langle 
\mathbf{k}_{12} , \mu_1, \mu_2 \Vert 
v_{12} \Vert \mathbf{k}_{12}',  \mu_1', \mu_2' \rangle
\label{bb.26a}
\end{equation} 
where 
\begin{equation}
k_{12}^{\mu} = B_c^{-1} (p_1+p_2)^{\mu}{}_{\nu} p_i^{\mu}
\label{bb.27} 
\end{equation}
is the two-body relative momentum and the magnetic quantum numbers are
associated with the two-body constituent spins
\begin{equation}
(0,\mathbf{j}_{2csi})^{\mu} = {1 \over m} (B^{-1}(k_{ij}) B^{-1}(p_i+p_j))
^{\mu}{}_{\nu} W_i^{\nu} .
\label{bb.28} 
\end{equation}

When these interactions are embedded in the three-body Hilbert space
the kernels (\ref{bb.26a}) are replaced by kernels that are
rotationally invariant functions of the three-body Jacobi momenta and
the three-body constituent spins.  In order to satisfy cluster
properties $\mathbf{k}_{ij}$ given by (\ref{bb.27}) is replaced by the
$\mathbf{k}_{ij}$ given by (\ref{bb.14}), the $p_i$ are replaced by
the corresponding $q_i$, and the two-body constituent spins
(\ref{bb.28}) are replaced by
\begin{equation}
(0,\mathbf{j}_{2(3)csi})^{\mu} = {1 \over m} (B^{-1}(k_{ij}) B^{-1}(q_i+q_j)
B^{-1}(P))
^{\mu}{}_{\nu} W_i^{\nu} .
\label{bb.29} 
\end{equation} 
These operators represent two-body constituent spins in the three-body
rest frame.  They agree with the two-body constituent spins
(\ref{bb.28}) that they replace in the three-body rest frame, but are
defined so they remain unchanged by canonical boosts out of the
three-body rest frame.  This ensures that they undergo the same Wigner
rotations as the kinematic three-body canonical spin under kinematic
Lorentz transformations.  Thus, the kernels (\ref{bb.26a}) 
are related by
\[
\langle \mathbf{P}, \mathbf{q}_i, \mathbf{k}_{jk}, \mu_1, \mu_2, \mu_3 
\vert v_{jk} 
\vert \mathbf{P}', \mathbf{q}'_i, \mathbf{k}'_{jk}, \mu_1', \mu_2', \mu_3' 
\rangle 
= 
\]
\[
\delta(\mathbf{P}-\mathbf{P}')\delta(\mathbf{q}_i -\mathbf{q}'_i) 
\delta_{\mu_i \mu_i'} \sum  
D^{1/2}_{\mu_j \bar{\mu}_j}[B_c(q_j)  B_c (q_j+q_k)B_c(k_{jk})] 
D^{1/2}_{\mu_k \bar{\mu}_k}[B_c(q_k)  B_c (q_j+q_k)B_c(-k_{jk})] 
\times
\]
\[
\langle \mathbf{k}_{jk} , \bar{\mu}_j, \bar{\mu}_k \Vert 
v_{jk} \Vert \mathbf{k}_{jk}',  \bar{\mu}_j', \bar{\mu}_k' \rangle
\times
\]
\begin{equation}
D^{1/2}_{\bar{\mu}_j' \mu_j'}[B_c^{-1}(k'_{jk}) B_c^{-1} (q'_j+q'_k)B_c^{-1}(q'_j)] 
D^{1/2}_{\bar{\mu}_k' \mu_k'}[B_c^{-1}(-k'_{jk}) B_c^{-1}(q'_j+q'_k)B_c^{-1}(q'_k)] .
\end{equation}
Here the unbarred magnetic quantum numbers are three-body constituent
spins while the barred magnetic quantum numbers are the two-body
constituent spins in the three-body rest frame.

Even though the spins in (\ref{bb.29}) transform the same way as the
three-body constituent spins, they differ from the three-body
constituent spins (\ref{bb.24}) by the Wigner rotation
\begin{equation}
(0,\mathbf{j}_{2(3)ics})^{\mu} = ((B^{-1}(k_{ij}) B^{-1}(q_i+q_i)
B^{-1}(q_i))^{\mu}{}_{\nu}  (0,\mathbf{j}_{3ics})^{\mu} .
\label{bb.30}
\end{equation}
When the two-body interactions are embedded in the three-body system
the spins are identified with the two-body constituent spins in the
three-body rest frame, as would be expected by cluster properties, but
in other frames they are defined to remain unchanged with respect to
canonical boosts out of the three-body rest frame.  The Wigner rotations
(\ref{b.10}) and (\ref{a2}) arise because the two-body subsystem is
moving in the three-body rest frame; however because the Wigner rotations
in (\ref{bb.30}) are functions of the $q_i$ rather than the $p_i$, 
both spins in (\ref{bb.30}) undergo the same Wigner rotations 
under kinematic Lorentz transformation.
Because of this it is also possible to construct the three-body
canonical spin using partial wave methods directly in a mixed
representation involving the barred spins in the interacting pair and
the unbarred spin for the spectator.  In the mixed representation 
the two-body interaction in the three-body Hilbert space has the 
simple form
\[
\langle \mathbf{P}, \mathbf{q}_i, 
\mathbf{k}_{jk}, \bar{\mu}_1, \bar{\mu}_2, \mu_3 
\vert v_{jk} 
\vert \mathbf{P}', \mathbf{q}'_i, \mathbf{k}'_{jk}, \bar{\mu}_1', 
\bar{\mu}_2', \mu_3' 
\rangle =
\]
\begin{equation}
\delta(\mathbf{P}-\mathbf{P}')\delta(\mathbf{q}_i -\mathbf{q}'_i) 
\delta_{\mu_i \mu_i'}   
\langle \mathbf{k}_{jk} , \bar{\mu}_j, \bar{\mu}_k \Vert 
v_{jk} \Vert \mathbf{k}_{jk}',  \bar{\mu}_j', \bar{\mu}_k' \rangle
\label{bb.30b}
\end{equation} 

For the two-body problem in the three-body Hilbert space it is
advantageous to use (\ref{bb.30b}) because spins (\ref{bb.29}) do not
require Wigner rotations.  However, with this choice each interacting
pair of particle must be treated using a permuted basis which
requires Wigner rotations in the permutation operators.  The
three-body forces are naturally expressed by a rotationally invariant
kernel in the three-body constituent spins.  When they are transformed
to a mixed basis that involves the spin (\ref{bb.29}) for one pair,
then it is necessary to transform two of the three-body constituent
spins with the Winger rotations in (\ref{bb.30}).  The calculations
performed in this work use a partial wave projection of the mixed
basis (\ref{bb.30b}), although the Wigner rotations in the three 
body-interaction are not yet included.

\section{Relativistic three-nucleon Faddeev equations with 
nucleon-nucleon forces}
\label{form}

The nucleon-deuteron scattering with neutron and protons interacting
through only a nucleon-nucleon interaction $v_{NN}$ is
described in terms of a breakup operator $T$ satisfying the
Faddeev-type integral equation~\cite{wit88,glo96}
\begin{eqnarray}
T\vert \phi \rangle  &=& t P \vert \phi \rangle + t P G_0 T \vert \phi 
\rangle .
\label{eq1a}
\end{eqnarray}
The two-nucleon $t$-matrix $t$ is the solution of the
Lippmann-Schwinger equation with the interaction
$v_{NN}$.  The permutation operator $P=P_{12}P_{23} +
P_{13}P_{23}$ is given in terms of the transposition operators,
$P_{ij}$, which interchanges nucleons i and j.  The incoming state $
\vert \phi \rangle = \vert \mathbf{q}_0 \rangle \vert \phi_d \rangle $
describes the free nucleon-deuteron motion with relative momentum
$\mathbf{q}_0$ and the deuteron state vector $\vert \phi_d \rangle$.
Finally $G_0$ is resolvent of the three-body center of mass kinetic
energy. Transition operators for the elastic nd scattering, $U$, and
breakup, $U_0$, are given in terms of $T$ by~\cite{wit88,glo96}
\begin{eqnarray}
U  &=& P G_0^{-1} + P T  ~, \cr
U_0 &=& (1+P)T ~.
\label{eq1c}
\end{eqnarray}

This is our standard nonrelativistic formulation, which is equivalent
to the nonrelativistic three-nucleon Schr\"odinger equation plus boundary
conditions.  The formal structure of these equations in the
relativistic case remains the same but the ingredients change. As
explained in ~\cite{relform} the relativistic three-nucleon rest Hamiltonian
(mass operator) has the same form as the nonrelativistic one, only
the momentum dependence of the kinetic energy and the relation
of the pair interactions in the three-body problem to the pair
interactions in the two-body problem change.  Consequently all the
formal steps leading to (\ref{eq1a}) and (\ref{eq1c}) remain the
same.

The free relativistic invariant mass of three identical nucleons of
 mass $m$ has the form ~\cite{witrel2} (see Eq.(\ref{bb.15}))
\begin{equation}
M_{0} = \sqrt{m_{230}^2 + {q}^2} + \sqrt{m^2 + {q}^2} 
\label{b.36}
\end{equation}
with spectator momentum $\mathbf{q}:=\mathbf{q}_1$ and 
the free two-body mass operator $m_{230}$ expressed in terms of
 the relative momentum $ \mathbf{k}:=\mathbf{k}_{23}$ in the $2-3$ center 
of momentum frame by (see Eq.(\ref{bb.16})) 
\begin{equation}
m_{230}\equiv 2\sqrt{{k}^2 + m^2} \equiv 2 \omega_m (k) ~.
\label{b.7}
\end{equation} 

As introduced in ~\cite{relform1} and in Eq.(\ref{bb.5}) 
the pair forces in the relativistic
three-nucleon $2+1$ mass operator are related to the two-body forces 
in the two-body problem, $v_{ij}$, by 
\begin{equation}
V_{(ij)(k)} = 
\sqrt{(m_{ij0} +v_{ij})^2 + {q}^2} - \sqrt{m_{ij0}^2 + {q}^2} ~,
\label{b.32}
\end{equation}
where $V= V({q}^2)$ reduces to the interaction
$v$ for $\mathbf{q} = 0$, which acts in the two-body
center of momentum frame. The momentum dependence
ensures that the resulting three-nucleon scattering
matrix satisfies space-like cluster properties in all
frames~\cite{relform1}.

The transition matrix $t$ that appears in the kernel of the Faddeev
equation (\ref{eq1a}) is obtained by solving the relativistic 
Lippmann-Schwinger equation as a function of $q^2$
\begin{eqnarray}
 t( \mathbf{k}, \mathbf{k}' ;  {q}^2) &=&
 V( \mathbf{k}, \mathbf{k}' ;  {q}^2) +
\int d^3k'' {
{V( \mathbf{k}, \mathbf{k}'' ; {q}^2)
 t( \mathbf{k}'', \mathbf{k}' ; {q}^2) }
\over{ \sqrt{ {( 2\omega_m( {{k'}}) )^{2} + {{q}}^{2}} }
-  \sqrt{ {(2\omega_m(  {{k''}}) )^{2}
+ {{q}}^{2}} } + i\epsilon  } } .
\label{eq2a}
\end{eqnarray}

The input two-body interaction $V$ is computed by solving the
nonlinear equation \cite{kamada1}
\begin{equation}
 \{ \sqrt{m_{ij0}^2 + {q}^2} , V_{(ij)(k)} \} + V_{(ij)(k)}^2 =
 4m v_{NN} ~,
\label{b.33}
\end{equation}
where  $v_{NN}$ is a nonrelativistic  nucleon-nucleon potential
fitted to the nucleon-nucleon data basis and where anticommutator 
 $\{ A , B \} \equiv AB +BA$. In case of $\mathbf{q}=\mathbf{0}$ that 
equation reduces to a nonlinear equation for the relativistic two-body 
interaction $v$. Therefore the problem of refitting all two-nucleon 
data when changing  from a nonrelativistic to a 
relativistic  Lippmann-Schwinger equation  is avoided. The  nonlinear 
equation (\ref{b.33}) can be solved by iteration \cite{kamada1}.
An alternative approach to determine 
 $t( \mathbf{k}, \mathbf{k}' ;  {q}^2)$ is described in \cite{CSP}.

The new relativistic ingredients in (\ref{eq1a})
and (\ref{eq1c}) will therefore
be the
$t$-operator (\ref{eq2a}) (expressed in partial waves) 
and the resolvent of the three-nucleon invariant mass
\begin{eqnarray}
G_0 &=& {{1}\over{E + i\epsilon - M_0}} ,
\label{eq2c}
\end{eqnarray}
with $M_0$ given by (\ref{b.36}).   
$E$ is the total three-nucleon invariant mass
expressed in terms of the initial
neutron momentum $\mathbf{q}_0$ relative to the deuteron by
\begin{eqnarray}
E &=& \sqrt{M_d^2 + {{q}}_0^{\ 2} } + \sqrt{m^2 + 
{{q}}_0^{\ 2}} ,
\label{eq2d}
\end{eqnarray}
with $M_d$  the deuteron rest mass. Related to the choice of the
permutation operator P the pair $i-j$ is chosen as $2-3$. 

Currently the Faddeev equation (\ref{eq1a}) in its nonrelativistic form is
numerically solved for any nucleon-nucleon interaction using a
momentum space partial-wave decomposition.  Details are presented in
\cite{wit88}.  Projecting (\ref{eq1a}) on such a basis turns it into a
coupled set of two-dimensional integral equations.  As shown
in~\cite{witrel1,witrel2}, in the relativistic case we can keep the
same formal structure, though the permutation operators are replaced
by the corresponding Racah coefficients for the Poincar\'e group which
include both Jacobians and Wigner rotations that do not appear in the
nonrelativistic permutation operators~\cite{glo96,book}.

In the nonrelativistic case the partial-wave projected momentum-space basis is
\begin{eqnarray}
 &~& \vert p q (ls)j
(\lambda{1\over{2}})IJ(t{1\over{2}})T \rangle ,
\label{eq2e}
\end{eqnarray}
where p and q are the magnitudes of standard Jacobi momenta (see
\cite{glo96,book}), obtained by transforming single particle momenta
to the rest frame of a two- or three-body system using a Galilean
boost, and $(ls)j$ are two-body quantum numbers with obvious meaning,
$(\lambda 1/2)I$ refer to the third, spectator nucleon, taken as the
nucleon $1$ and described by the momentum q, $J$ is the total
three-nucleon angular momentum and the rest are isospin quantum
numbers.  In the relativistic case this basis is replaced by the
Poincar\'e irreducible states defined as~\cite{witrel2}
\begin{eqnarray}
&& \langle \mathbf{p}_1, \mu_1' , \mathbf{p}_2, \mu_2' ,\mathbf{p}_3, \mu_3'
\vert  (J, q) \mathbf{P}=\mathbf{0} ,\mu ; 
\lambda ,I ,j_{23} ,k_{23}, l_{23} ,s_{23} \rangle ~
=
\cr
&&
~ \delta( \mathbf{0}-  \mathbf{q}_1 -\mathbf{q}_2
- \mathbf{q}_3 )
{1\over{ N({q}_2, {q}_3) } }  { \delta(q_1 -q ) \over { q^2 } }
{ \delta(k(~ \mathbf{q}_2,\mathbf{q}_3~ )-k  )
\over { k^2  } } ~ \cr
&&
 \sum_{\mu_2 \mu_3 \mu_s }
 \sum_{\mu_l \mu_{\lambda} \mu_I }
  ( {1\over{2}}, \mu_2 ,{1\over{2}}, \mu_3 \vert s, \mu_s )
 (  l,  \mu_l, s, \mu_s, \vert j, \mu_j )
(\lambda, \mu_{\lambda}, {1\over{2}}, \mu'_1 \vert I, \mu_I )
( j, \mu_j, I, \mu_I \vert J, \mu) \cr
&& 
 Y_{\lambda \mu_{\lambda}}(\hat{\mathbf{q}}_1)
 Y_{l \mu_l}(\hat{\mathbf{k}}(\mathbf{q}_2,\mathbf{q}_3 ) )  
D^{1\over{2} }_{\mu'_2 \mu_2}
[R_{wc}(B_c(-q_1),{k}_2(\mathbf{q}_2,\mathbf{q}_3 ))] 
D^{1\over{2} }_{\mu_3' \mu_3}
[R_{wc}(B_c(-q_1),
{k}_3(\mathbf{q}_2,\mathbf{q}_3 ))] ~,
\label{b.10}
\end{eqnarray}
where $N({q}_2, {q}_3)$ is given by (\ref{a2_a}) in Appendix
\ref{app_a} and $\vec k( \vec q_2,\vec q_3)$ by (\ref{equ.B1}) in 
Appendix \ref{app_b}.
These states are labeled by the same quantum
numbers as the corresponding non-relativistic basis states.

The basis states (\ref{b.10}) are used for the evaluation of the
partial wave representation of the permutation operator P with Wigner
rotations of spin states for nucleons $2$ and $3$ included.  In the
relativistic case we adopt the following short-hand notation for the
Poincar\'e irreducible three-body states, which also includes isospin quantum
numbers coupled in the same order:
\begin{equation}
\vert k, q ,\alpha \rangle :=
\vert k q (ls)j (\lambda ,{1\over{2}}) IJ(t{1\over {2}} )T \rangle =
\vert (J, q) \mathbf{P}=\mathbf{0}
,\mu ; \lambda ,I ,j_{23} ,k_{23}, l_{23} ,s_{23}
\rangle \vert  (t{1\over {2}} ) T \rangle ~.
\label{b.13}
\end{equation}
Equipped with that, projecting  (\ref{eq1a}) onto the basis states
$\vert k, q ,\alpha \rangle$
one encounters, using the nonrelativistic notation of Ref.~\cite{book}
\begin{eqnarray}
_1\langle k q \alpha \vert P \vert k' q' \alpha' \rangle_1
&=&  _1\langle k q \alpha \vert k' q' \alpha' \rangle_2 +
  _1\langle k q \alpha \vert k' q' \alpha' \rangle_3
 =  2~  _1\langle k q \alpha \vert k' q' \alpha' \rangle_2 .
\label{eqm6}
\end{eqnarray}
This is evaluated by inserting the complete basis of 
single-particle states $\vert
\mathbf{p}_1, \mu_1 ,\mathbf{p}_2, \mu_2, \mathbf{p}_3 ,\mu_3 \rangle$ and
using (\ref{b.10}).   It can
be expressed in a form which resembles closely the corresponding 
non-relativistic expression ~\cite{book,glo96}
\begin{eqnarray}
_1\langle k ~ q ~ \alpha \vert ~ P ~ \vert k' ~ q' ~ \alpha' \rangle_1 &=&
\int_{-1}^{1} dx { {\delta(k-\pi_1)} \over { k^{2} }  } ~
{ {\delta(k'-\pi_2)} \over { k'^{2} }  } ~  \cr
&~&
{1\over {N_1(q, q',x) } } ~ {1\over {N_2(q, q',x) } } ~
G_{\alpha \alpha'}^{BB} (q, q', x) ,
\label{eqm7}
\end{eqnarray}
where all ingredients are defined in Appendix \ref{a2} of
Ref.~\cite{witrel2}.  The rotational invariance of the nucleon-nucleon
interaction in this basis ensures that all three nucleon-nucleon
interactions commute with the spin Casimir operator of the
non-interacting three-nucleon system.  This allows the interactions to
be added in a manner that preserves the underlying Poincar\'e
symmetry.

Due to the short-range nature of the nucleon-nucleon interaction it can be
considered negligible beyond a certain value $j_{max}$ of the total
angular momentum in the two-nucleon subsystem. Generally with
increasing energy $j_{max}$ will also increase. For $j > j_{max}$ we
set the t-matrix to zero, which yields a finite number of coupled
channels for each total angular momentum J and total parity
$\pi=(-)^{l+\lambda}$ of the three-nucleon system.  To achieve
converged results at incoming nucleon laboratory energies below
$\approx 250$~MeV all partial wave states with total angular momenta
of the two-nucleon subsystem up to $j_{max}=5$ and all total angular
momenta of the three-nucleon system up to $J=25/2$ must be taken into
account. This
leads to a system of up to 143 coupled integral equations in two
continuous variables for a given $J$ and parity.  For the details of
 the numerical performance we refer to \cite{glo96,book,witrel1}.  The
solution of these equations can be used to construct an exactly
Poincar\'e invariant scattering operator.

\section{Relativistic three nucleon Faddeev equations 
with a three-nucleon force included}
\label{3nfmatrixelement}

In the standard nonrelativistic formulation when in addition to
pairwise interactions $v_{NN}$ between three nucleons
also a three-nucleon force is included, a new term $V_4$ appears in a
potential energy of the three-nucleon system
\begin{equation}
\label{eqU2}
V_4 = V_4^{(1)} + V_4^{(2)} + V_4^{(3)}~.
\end{equation}
Each $V_4^{(i)}$ is symmetric under  exchange of the nucleons
$j$ and $k$ ($i, j, k=1,2,3$ and $j\ne i \ne k$). 
In the $2\pi$-exchange three-nucleon force $V_4^{(1)}$ is
a contribution to the three-nucleon potential from (off-shell) rescattering
of a pion on nucleon 1. 

When a three-nucleon force is acting then on top of rescatterings
among three nucleons induced by pairwise forces only, which are summed
up in integral equation (\ref{eq1a}), additional rescatterings induced
by three-nucleon force and nucleon-nucleon force appear.

Therefore  Faddeev equation (\ref{eq1a}) changes to
\begin{equation}
\label{eqU3}
T \vert \phi \rangle  = \ t \, P \, \vert \phi \rangle  \ 
+ \ ( 1 + t G_0 ) \, V_4^{(1)} \, ( 1 + P ) \, \vert \phi \rangle  \
+  \ t \, P \, G_0 \, T \vert \phi \rangle \
+ \ ( 1 + t G_0 ) \, V_4^{(1)} \, ( 1 + P ) \, G_0 \, T \vert \phi \rangle ~,
\end{equation}
with one new contribution in the leading term and in the
kernel~\cite{glo96,hub97}.  While the breakup transition operator
$U_0$ preserves its form (\ref{eq1c}), in the elastic scattering
operator $U$ two new contributions appear~\cite{glo96,hub97}
\begin{equation}
 U  = P G_0^{-1} \ + \ V_4^{(1)} \, ( 1 + P ) \ + \ 
\ P T \  + \
V_4^{(1)} \, ( 1 + P ) \, G_0 \, T  ~.
\label{eqU1}
\end{equation}
The second term is due to a single interaction of three-nucleons via a
three-nucleon force and the fourth results from rescattering among
three nucleons induced by two- and three-nucleon forces with a
three-nucleon force as the final interaction.

After projecting on a partial-wave momentum-space basis equation
(\ref{eqU3}) becomes a system of 2-dimensional coupled integral
equations which can be solved numerically exactly for any nuclear
force.  Since the three-nucleon force is short-ranged its inclusion needs
to be carried through only for all total angular momenta of the three
nucleon system up to $J=13/2$.  As mentioned in section \ref{form},
the longer ranged two-nucleon interactions require states up to
$J=25/2$.  For details of the formalism and numerical performance in
case of the nonrelativistic formulation when three-nucleon force is
included we refer to Refs.~\cite{glo96,wit88,hub93}.

For relativistic calculations without a three-nucleon force, briefly
described in previous section, the details of the numerical treatment
are given in \cite{witrel1,witrel2}.  When a three-nucleon force is
added two new terms in (\ref{eqU3})
 contain the free three-nucleon propagator $G_0$. Since in the basis
$|\mathbf{k}, \mathbf{q} >$ (see Appendix \ref{app_b}) 
 the three-nucleon invariant mass $M_0$ is
diagonal, $G_0$ is given by
\begin{equation}
 \langle \mathbf{k}, \mathbf{q} | G_0 | \mathbf{k}', \mathbf{q}' \rangle  = 
\delta( \mathbf{k} - \mathbf{k}') \delta( \mathbf{q} - \mathbf{q}') 
\frac {1}{E-\sqrt{m^2+q^2} - \sqrt{4(k^2+m^2) + q^2} + i\epsilon } ~. 
\end{equation}
That means that performing integrations over momenta $k'$ and $q'$ in
the intermediate states $|k', q', \alpha' \rangle$ during the calculation of
matrix elements for these new terms, the simple pole singularity
occurs for momenta $q' < q_{max}$ at $k' = k_0$, where $q_{max}$ is
given by the total three-nucleon center of momentum energy $E$ 
through $E=\sqrt{4m^2 + q_{max}^2} + \sqrt{m^2+q_{max}^2}$.

For a given $q'$-value the momentum $k_0$ is the solution of
$E=\sqrt{4(m^2+k_0^2) + {q'}^2} + \sqrt{m^2+{q'}^2}$. The treatment of
that pole, as well as of the deuteron bound state pole in $T$, which
occurs at $q'=q_0$ for channels $\alpha'$ containing the deuteron
quantum numbers, was done using subtraction method \cite{wit88,glo96}.

The nonrelativistic treatment of (\ref{eqU3}) requires matrix elements
of $V_4^{(1)}(1+P)$ calculated in a partial-wave basis with standard
Jacobi momenta: $\langle p, q, \alpha |V_4^{(1)}(1+P)| p', q', \alpha'
\rangle$.  In the relativistic calculations, however, one needs them
in the new, relativistic basis $| k, q, \alpha \rangle$.  The
generation of three-nucleon force partial-wave matrix elements is the
most time consuming part of three-nucleon continuum Faddeev
calculations.  One way to reduce the computer time is to perform a
transformation of the existing, standard Jacobi momenta matrix
elements to the relativistic basis.  In Appendix \ref{app_a} we give
the expression (\ref{a10}) for such transformation which is valid in
the general case, when in addition to boost also Wigner spin rotations
are taken into account.  The complex structure of that transformation,
where in addition to the summation over numerous intermediate states
with geometrical coefficients, also involved are two integrations and two
interpolations over the momenta $p$ and $p'$, prevents, due to the
large amount of computing time and computer resources required, the
application of that transformation in fully converged calculations.

It seems thus unavoidable that in order to get matrix elements
$\langle k, q, \alpha |V_4^{(1)}| k', q', \alpha'\rangle$ one must
start from a commonly given expression for a three-nucleon force in
terms of individual nucleons momenta and their spin and isospin
operators and to apply to that expression the recently proposed
automatized partial wave decomposition \cite{automa1,automa2}. To that
aim we derived in Appendix \ref{app_b} relation (\ref{transf_1}) which
allows to express matrix element of a three-nucleon force in a
3-dimensional relativistic basis $\langle \mathbf{k}, \mathbf{q}
|V_4^{(1)}| \mathbf{k}', \mathbf{q}' \rangle$ by its matrix element in
the individual nucleons momenta basis.  In this basis $\mathbf{q}$ and
$\mathbf{k}$ undergo identical Wigner rotations under kinematic boost
of the three-nucleon system.  The nucleon spins are defined to be the
three-body constituent spins (canonical spins measured by using a
rotationless boost to the three-body center of momentum frame).  To
use them in the $| k, q, \alpha\rangle$ basis the spins for the
$\alpha$ pair must be Wigner rotated before they are coupled.  The
alternative is to use the representation where three spins are
three-body constituent spins; in this case all three of the two-body
interactions will have Wigner rotations that convert the two-body
constituent spins in the three body rest frame to three-body
constituent spins.  The three-nucleon force will have no Wigner
rotations.  In this representation all of the spins can be coupled
using standard partial wave methods.  For our calculations we work in
the $| k, q, \alpha\rangle$ basis, but do not account for the Wigner
rotations in the three-nucleon interaction for the reasons discussed in
the previous paragraph.  This allows us to treat the spins in the
three-nucleon force using conventional methods.  This assumes the Wigner
rotations in the three-nucleon force can be neglected.  Neglecting these
Wigner rotations has no effect on the relativistic invariance or 
$S$-matrix cluster properties. 

\section{Results}
\label{results}

To study the importance of a consistent treatment of both relativity
and a three-nucleon force we numerically solved the three-nucleon
Faddeev equations for neutron-deuteron scattering at the neutron
laboratory energies $E_n^{lab}=70$, $135$, $200$ and $250$~MeV. As
dynamical input we took the nonrelativistic nucleon-nucleon potential
CD~Bonn~\cite{CDBOnucleon-nucleon} and TM99 three-nucleon
force~\cite{TM,TMgl,TM99}.  The cut-off parameter $\Lambda$ of that
three-nucleon force was adjusted to $\Lambda=4.469$ in units of the
pion mass, $m_{\pi}$, to give, together with the CD~Bonn potential,
the experimental binding energy of $^3$H.  At each energy we generated
solutions of nonrelativistic and relativistic three-nucleon Faddeev
equation, without and with TM99 three-nucleon force included.  For
relativistic case we produced, starting from the CD~Bonn potential and
solving nonlinear equation (\ref{b.33}) at the required spectator nucleon
momenta $q$, the relativistic, on-shell equivalent interaction with
boost effects incorporated exactly.  That interaction served as
dynamical input to calculate, using the relativistic Lippmann-Schwinger
equation (\ref{eq2a}) the relativistic off-shell t-matrix $t$ that
appears in Faddeev equations.

Since in \cite{witrel2} it was found that effects of Wigner spin
rotations are practically negligible in the studied energy range, we
neglected them in the present study. When performing relativistic
calculations with three-nucleon force included one requires matrix
elements of the TM99 three-nucleon force in a relativistic momentum
space basis, where the relative momentum of two nucleons in their
c.m. system, $\mathbf{k}$, replaces standard Jacobi momentum
$\mathbf{p}$. That momentum $\mathbf{k}$ together with spectator
nucleon momentum $\mathbf{q}$, equal in magnitude and opposite to the
total momentum of the free pair in three-nucleon center of momentum system,
unambiguously define the configuration of three nucleons.  Since it is
the region of small and not large momenta which is most important when
solving Faddeev equations it seems reasonable to assume that the
momenta $\mathbf{k}$ and $\mathbf{p}$ do not differ
substantially. Therefore, in order to avoid calculations of the TM99
three-nucleon force matrix elements in a relativistic basis $|k, q,
\alpha \rangle$ we assumed, that the matrix elements in a relativistic
and nonrelativistic bases are equal:
\begin{eqnarray}
\langle k, q, \alpha \vert V_4^{(1)} \vert k', q', \alpha' \rangle
&=&  \langle p=k, q, \alpha  \vert V_4^{(1)} \vert p'=k', q', \alpha' 
\rangle ~.
\label{eq_base1}
\end{eqnarray}
That assumption allowed us to use the existing matrix elements of the
TM99 three-nucleon force. 

\begin{figure}
\includegraphics[scale=0.7]{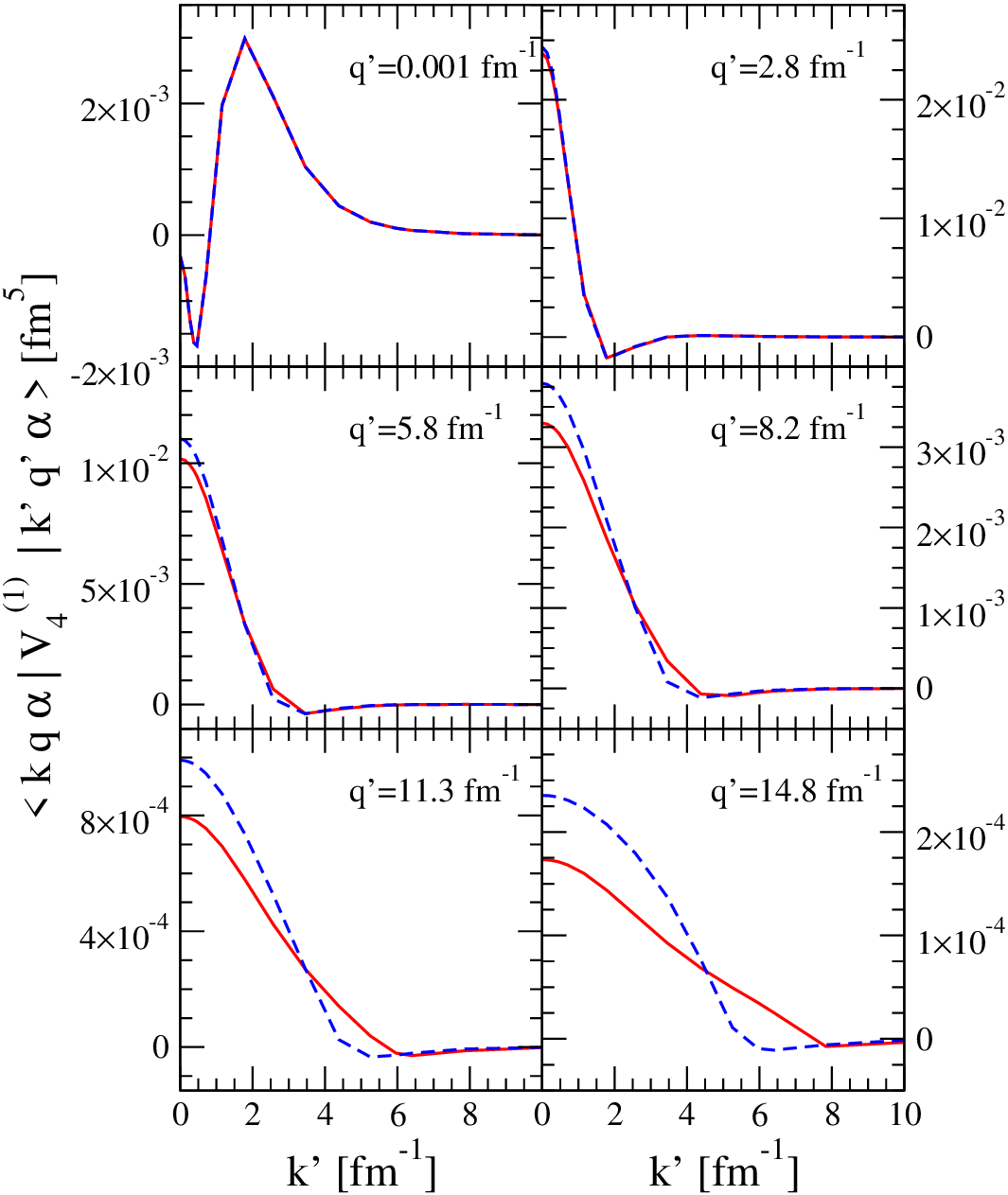}
\caption{
(color online) 
The matrix element of the TM99 3NF  in relativistic-  
 ($\langle k, q, \alpha \vert V_4^{(1)} \vert k', q', \alpha \rangle$
- blue dashed line) and nonrelativistic-basis 
 ($\langle p, q, \alpha \vert V_4^{(1)} \vert p', q', \alpha \rangle$
- red solid line) for the total angular momentum and parity 
of the 3N system $J^{\pi} = {\frac {1} {2}}^+$ and channel $\alpha = 
| (00)0 (0\frac {1} {2})\frac {1} {2} (1\frac {1} {2})\frac {1} {2}>$.
   The momenta $p=k=0.132$~fm$^{-1}$ and  $q=0.132$~fm$^{-1}$.
}
\label{fig0a_1}
\end{figure}

\begin{figure}
\includegraphics[scale=0.7]{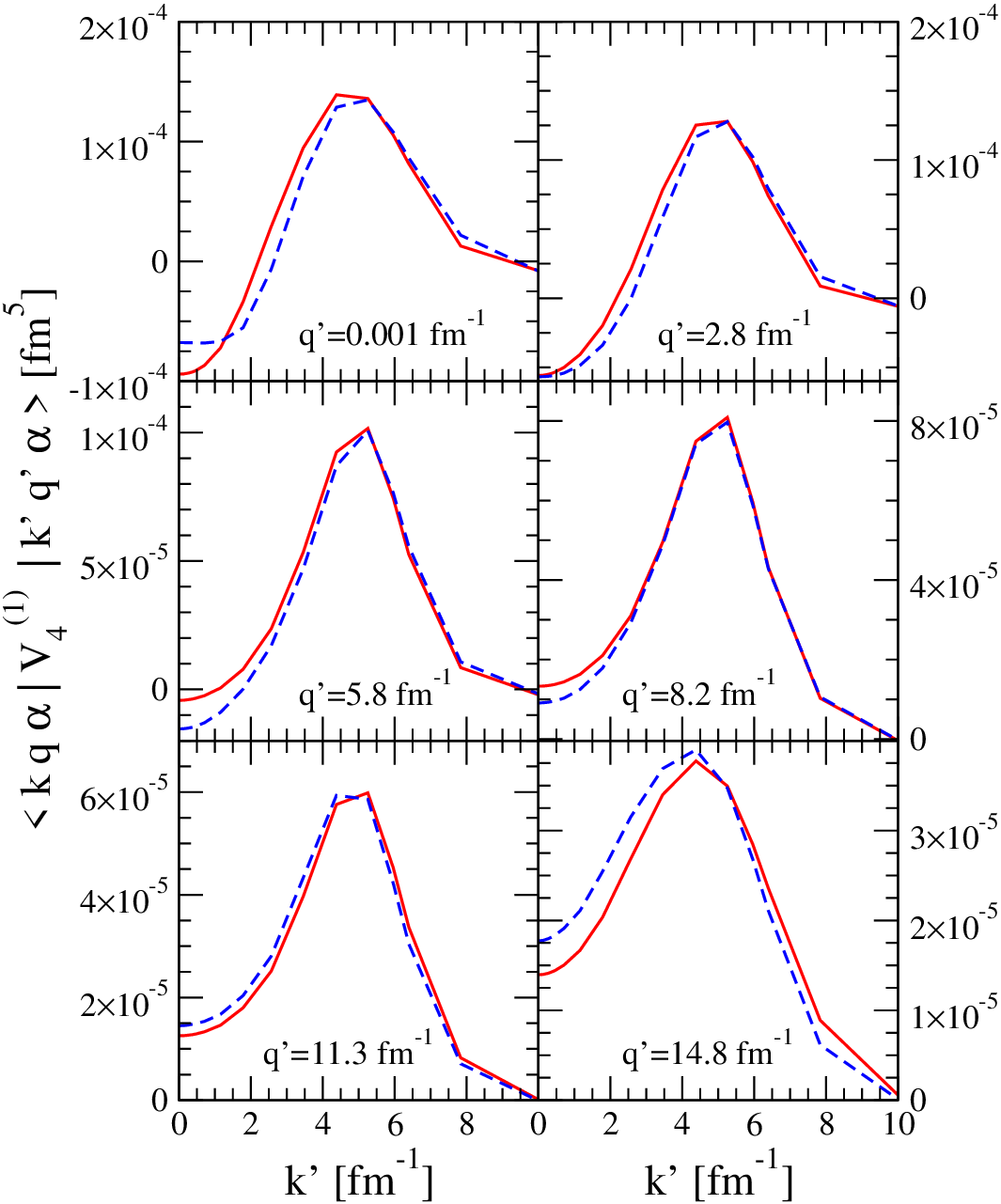}
\caption{
(color online) 
The same as in Fig.~\ref{fig0a_1} 
but for the  
momenta $p=k=5.25$~fm$^{-1}$ and  
 $q=8.24$~fm$^{-1}$.
}
\label{fig0a_2}
\end{figure}

To check quality of  the approximation (\ref{eq_base1}) we compared 
 the matrix element of the
 TM99 3NF in the relativistic basis,  
$\langle k, q, \alpha \vert V_4^{(1)} \vert k', q', \alpha' \rangle$, 
calculated according to (\ref{transf_1}) and using automatized partial wave
expansion of Ref.~\cite{automa2} (what corresponds to neglection of
Wigner spin rotations in (\ref{transf_2})), 
with  the corresponding matrix element in the
standard, nonrelativistic basis,  
$\langle p, q, \alpha \vert V_4^{(1)} \vert p', q', \alpha' \rangle$, at a
number of the spectator momentum values. 
In Figs.~\ref{fig0a_1} and \ref{fig0a_2} we exemplify the typical
behavior showing 
at a number of  q' values and at a fixed $p=k$, taking  two different
values of $q$,  
 the $k'(=p')$ dependence of  
these matrix elements for a particular channel $\alpha =
\alpha' = |(00)0 (0\frac {1} {2})\frac {1} {2} 
(1\frac {1} {2})\frac {1} {2} >$. 
 As expected,  clear differences between these matrix elements occur
 only at very large values of the 
 spectator momentum q, where magnitudes of these matrix elements
 are small. 
This justifies application of the approximation (\ref{eq_base1}) in the present
study. 

\begin{figure}
\includegraphics[scale=0.7]{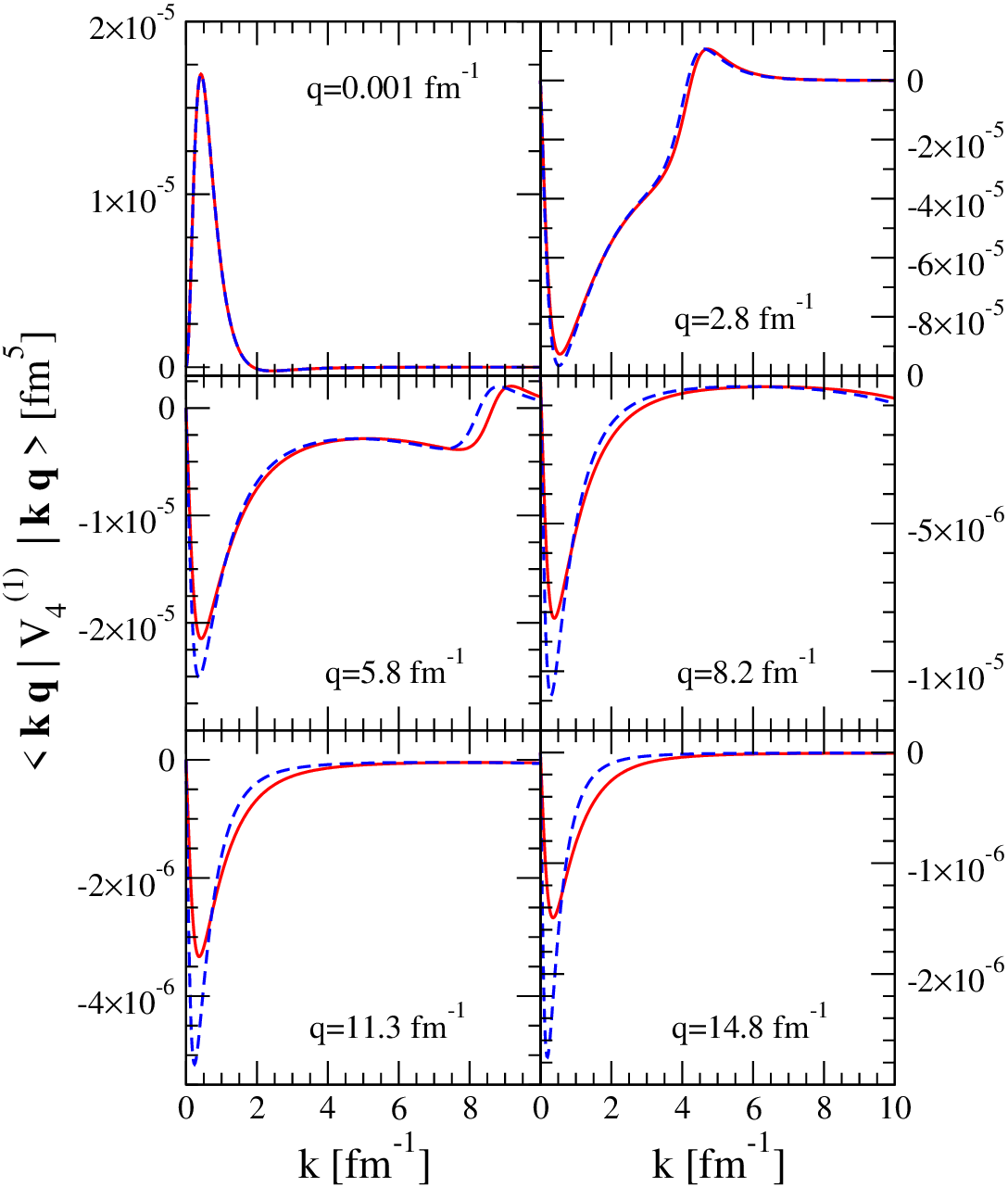}
\caption{
(color online) 
The matrix element of the TM99 3NF  in relativistic- 
 ($\langle \mathbf{k}, \mathbf{q} \vert V_4^{(1)} \vert \mathbf{k},
  \mathbf{q} \rangle$
- blue dashed line) and nonrelativistic-basis 
 ($\langle \mathbf{p}, \mathbf{q} \vert V_4^{(1)} \vert \mathbf{p},
\mathbf{q} \rangle$
- red solid line). 
   They are shown as a function of $k$ at a number of $q$ values assuming
   that all  momenta are  parallel to the unit vector 
$ \left( \frac1{\sqrt{3}}, \frac1{\sqrt{3}}, \frac1{\sqrt{3}}\,
   \right) $, all spin magnetic quantum numbers 
 in the initial and final state are equal $\frac {1}
 {2}$, and isospins  $t=t'=0$.
}
\label{fig0b}
\end{figure}

The approximation (\ref{eq_base1}) can be investigated also directly for 
the three-dimensional 
matrix elements, comparing 
$\left\langle {\mathbf{k}, \mathbf{q}}~ \right|V_4^{(1)}
\left| {\mathbf{k}'}, \mathbf{q}' \right\rangle$ 
and 
$\left\langle {\mathbf{p}, \mathbf{q}}~ \right|V_4^{(1)}
\left| {\mathbf{p}'}, \mathbf{q}' \right\rangle$.
The connection between these matrix elements 
is given by  (\ref{transf_1}) in Appendix \ref{app_b}. 
 They depend on momentum vectors and spin-isospin quantum numbers 
in the initial and final state. In Fig.~\ref{fig0b} we show a particularly 
simple case, where $t=t'=0$, all four momenta are parallel 
to the unit vector 
$ \left( \frac1{\sqrt{3}}, \frac1{\sqrt{3}}, \frac1{\sqrt{3}}\, \right) $
and all spin magnetic quantum numbers are equal $\frac12$.
We display
$\left\langle {\mathbf{k}, \mathbf{q}}~ \right|V_4^{(1)}
\left| {\mathbf{k}}, \mathbf{q} \right\rangle$
and
$\left\langle {\mathbf{p}, \mathbf{q}}~ \right|V_4^{(1)}
\left| {\mathbf{p}}, \mathbf{q} \right\rangle$
for several $q$ values as a function of  
$k$. We see how the difference develops gradually with increasing $q$,
resembling the picture seen for partial wave decomposed matrix elements.

Transition amplitudes for elastic neutron-deuteron scattering and
breakup based on that set of four solutions of three-nucleon Faddeev
equations, are used to predict numerous observables for both
reactions. By comparing these observables conclusions on how strongly
three-nucleon force effects depend on relativity were drawn. In the
following subsections we show and discuss results for the cross
section and numerous spin observables, separately for elastic
scattering and breakup reactions.

\subsection{Elastic scattering}
\label{elastic}

\begin{figure}
\includegraphics[scale=0.9]{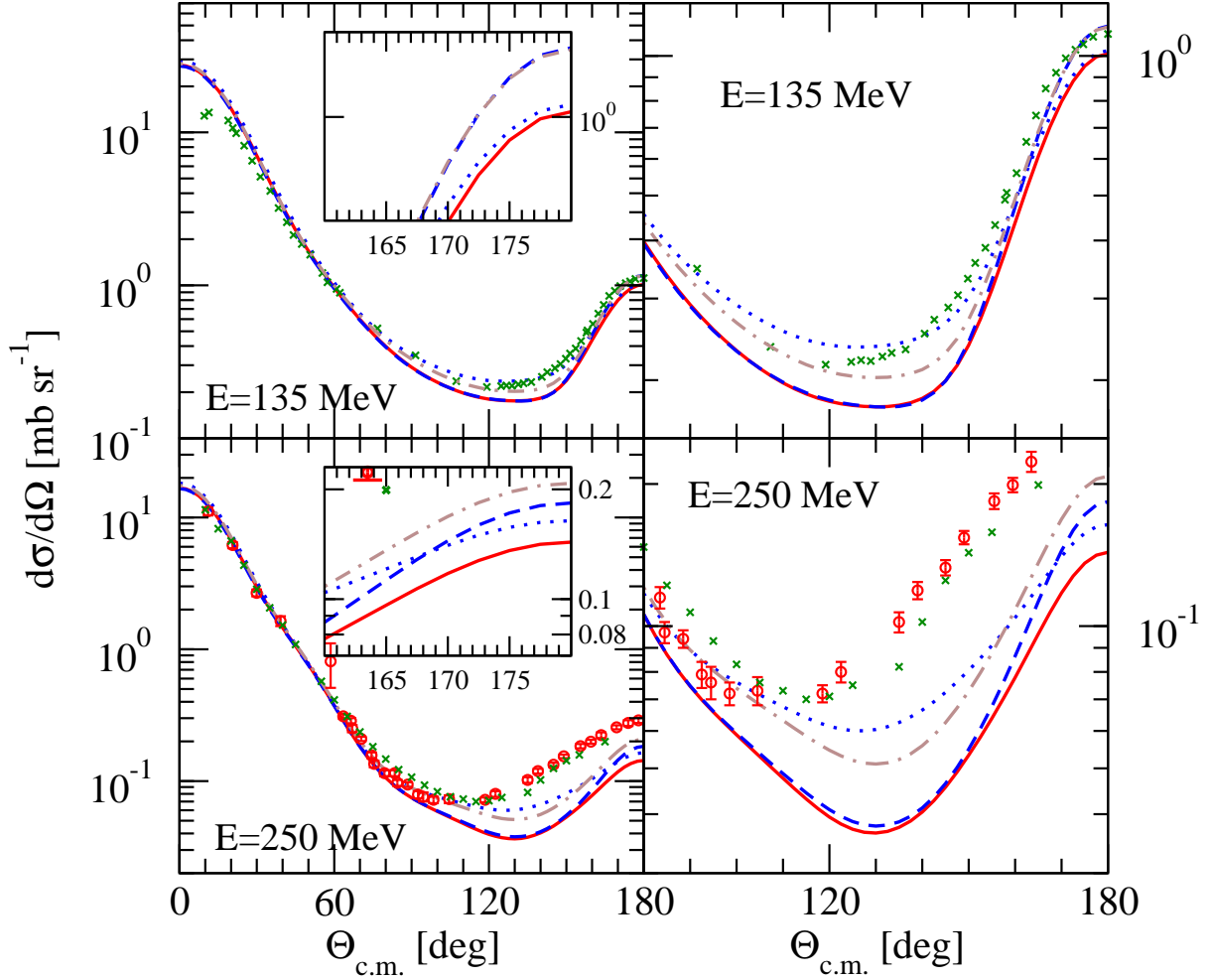}
\caption{
(color online) 
The elastic nd scattering angular distributions at the incoming
neutron lab.  
energy $E=135$ and $250$~MeV. 
The solid (red) and dotted (blue) lines are  results 
of the non-relativistic Faddeev calculations with
the CD~Bonn potential alone and combined with TM99 three-nucleon
force, 
respectively. 
The relativistic predictions based on CD~Bonn potential without 
Wigner spin rotations are shown by the dashed (blue) lines. The
dashed-dotted (brown) lines show results of relativistic calculations
with the TM99 three-nucleon force included. 
The pd data (x-es) at $135$~MeV are from ref.~\cite{sek02} 
and at $250$~MeV  from ref.~\cite{hat02}. At  $250$~MeV also nd data of
ref.~\cite{maedand} are shown by circles. The inserts 
and figures in the right column display details of the cross
sections in specific angular ranges. 
}
\label{fig1}
\end{figure}

At higher energies of the incoming nucleon three-nucleon forces play
significant role in determining the angular distribution of the
elastic neutron deuteron scattering. The clear
evidence of three-body force
effects start to
develop at $E_n^{lab} \approx 65$~MeV for scattering
angles close to a minimum of the cross section, which at $65$~MeV
occurs at $\theta_{c.m.} \approx 105^o$ \cite{wit98,wit01}. With
increasing energy of the three-nucleon system not only the magnitude
of predicted three-nucleon force effect increases but it also
influences the cross section in a wider range of angles, which at
$250$~MeV covers $90^o \le \theta_{c.m.} \le 180^o$
\cite{wit98,wit01}.  The standard $2\pi$-exchange three-nucleon
forces, such as TM99~\cite{TM99} or Urbana IX~\cite{uIX}, are able to
account for existing discrepancies between theoretical cross sections
obtained with realistic nucleon-nucleon potentials and data only up to
$E_n^{lab} \approx 135$~MeV. Data at larger energies in a region
of angles ranging from the cross section minimum up to $180^o$ are
drastically underestimated even when $2\pi$-exchange three-nucleon
forces are included in the calculations. This is exemplified on
Fig.~\ref{fig1}, where solid (red) lines are nonrelativistic
predictions based on the CD~Bonn potential alone and dotted (blue)
lines are results obtained when the CD~Bonn potential was combined
with the TM99 three-nucleon force.

Since effects of relativity for predictions based on two-nucleon
forces only are restricted to very backward angles $\theta_{c.m.} \ge
160^o$ \cite{witrel1} (see also Fig.~\ref{fig1} where dashed (blue)
lines are relativistic predictions based on the CD~Bonn potential),
the drastic discrepancy between data and theory seen at $250$~MeV
would indicate that at such large energies shorter-ranged
three-nucleon force components, not taken into account in these
calculations, start to play significant role. The possibility, that
including such three-nucleon force contributions would indeed help to
improve description of the cross section data is further supported by
an interesting pattern revealed when the TM99 three-nucleon force is
included into relativistic calculations.  Namely, when a consistent
treatment of relativity and a three-nucleon force as described in the
present study is made, then the resulting changes of the cross section
are not a simple incoherent sum of effects due to relativity, seen
when two-nucleon forces alone are acting, and three-nucleon force
effects found in nonrelativistic calculations. The relativity
modulates effects exerted by the TM99 three-nucleon force on the cross
section found in nonrelativistic calculations and the magnitude of
this modulation depends from the scattering angle.  While at backward
angles the nonrelativistic cross section with a three-nucleon force
included is further enhanced by relativity, in a region of center of momentum 
angles near the cross section minimum the magnitude of three-nucleon force
effects seen in nonrelativistic calculations is strongly reduced by
relativity (dashed-dotted (brown) lines in Fig.~\ref{fig1}).

Also  elastic scattering polarization observables reveal such
incoherent and angle-dependent modulation of three-nucleon force effects by
relativity. The details, however,  depend on the particular spin
observable under study and every conceivable scenario can be found. 

\begin{figure}
\includegraphics[scale=0.8]{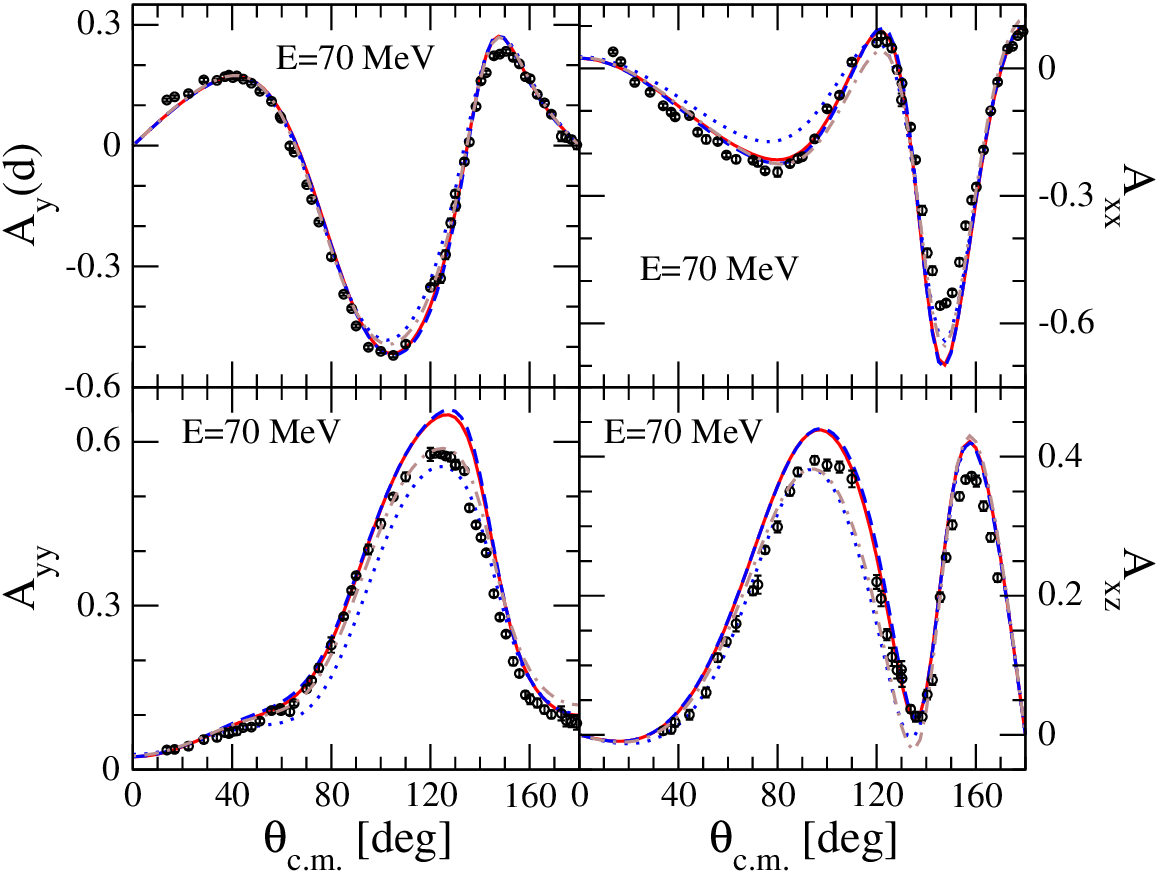}
\caption{
(color online)
The vector (deuteron) $A_y(d)$ and tensor analyzing powers 
$A_{xx}$, $A_{yy}$, and $A_{xz}$  in elastic nd scattering at the incoming
neutron lab. energy $E=70$~MeV.
For description of lines see Fig.\ref{fig1}. 
The pd  data (open circles) are from \cite{sek02}.
}
\label{fig2}
\end{figure}

\begin{figure}
\includegraphics[scale=0.8]{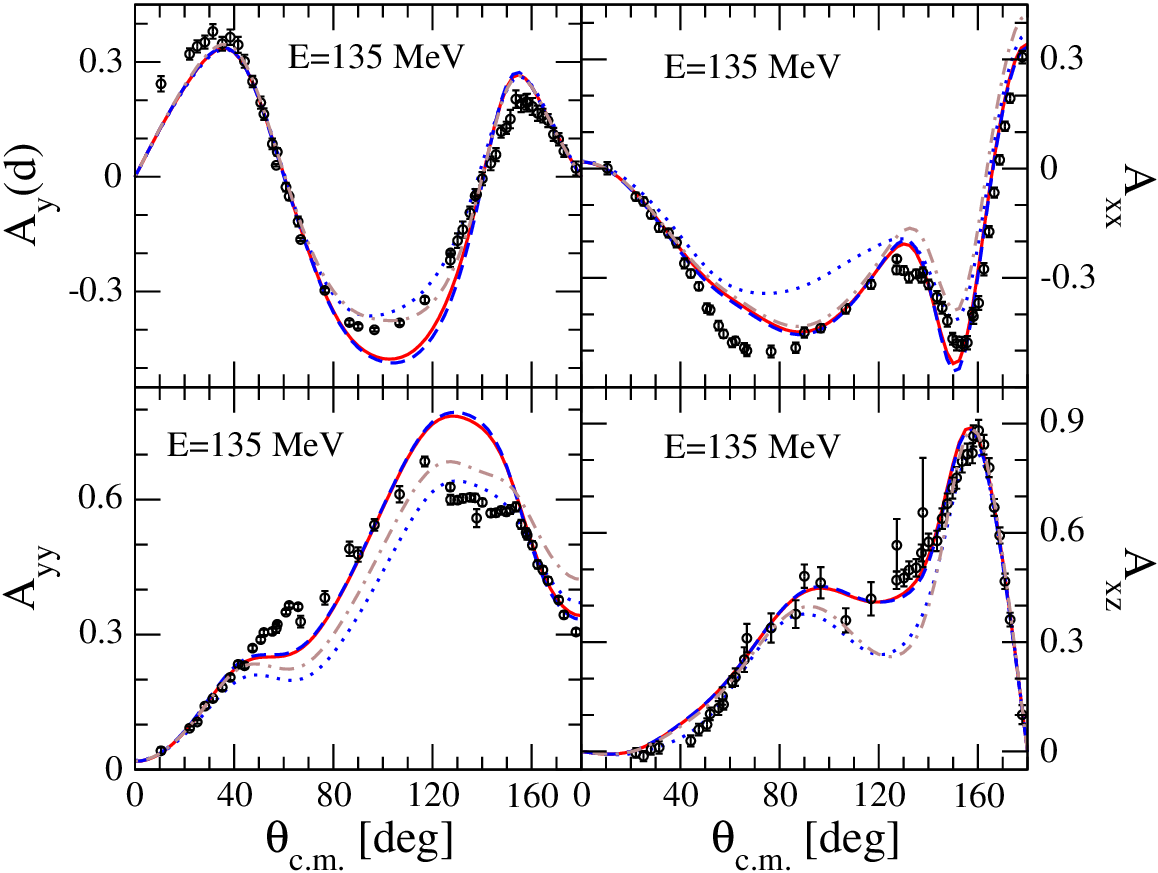}
\caption{
(color online)
The vector (deuteron) $A_y(d)$ and tensor analyzing powers 
$A_{xx}$, $A_{yy}$, and
$A_{xz}$  in elastic nd scattering at the incoming
neutron lab. energy $E=135$~MeV.
For description of lines see Fig.\ref{fig1}. 
The pd  data (open circles) are from \cite{sek02}.
}
\label{fig3}
\end{figure}

\begin{figure}
\includegraphics[scale=0.8]{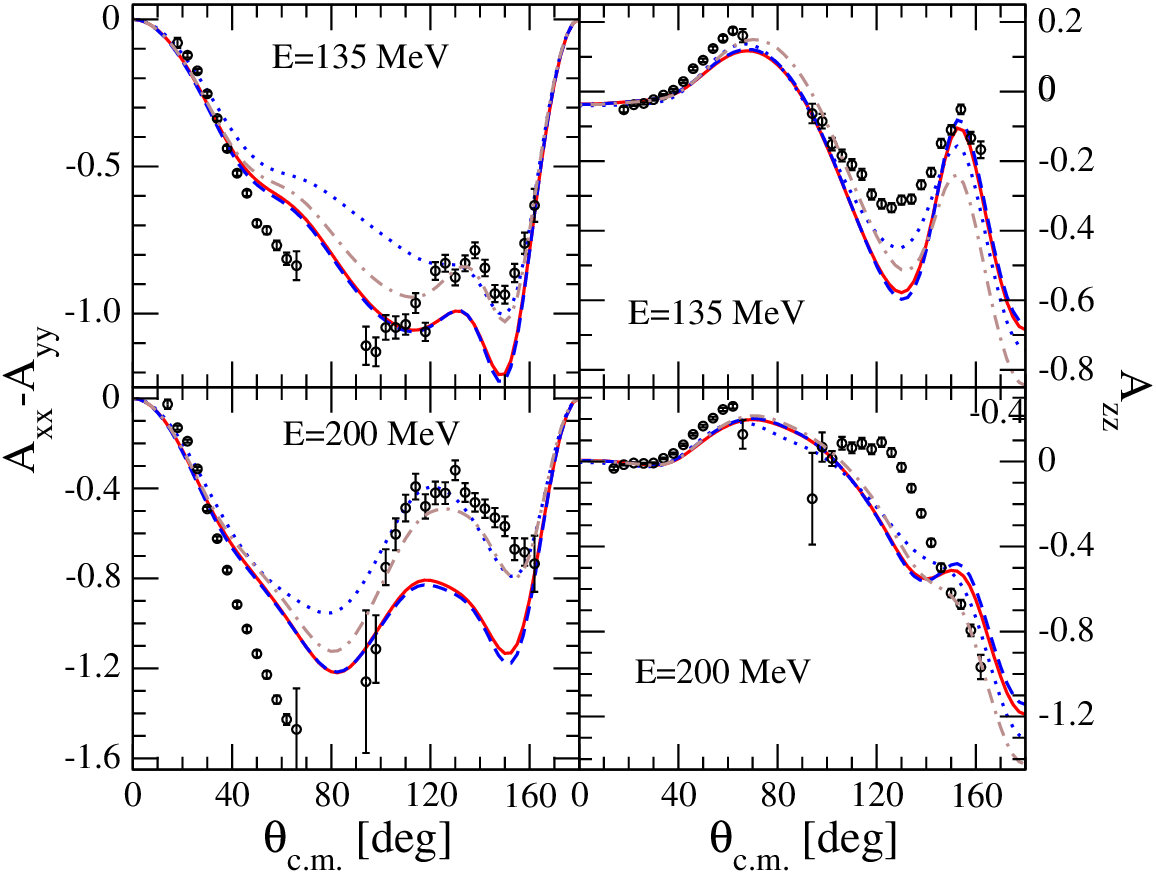}
\caption{
(color online)
The tensor analyzing powers  $A_{xx} - A_{yy}$ and
$A_{zz}$ in elastic nd scattering at the incoming neutron lab.  
energy $E=135$ and $E=200$~MeV.
For description of lines see Fig.\ref{fig1}. 
The pd  data (open circles) are from \cite{indiana}.
}
\label{fig4}
\end{figure}

\begin{figure}
\includegraphics[scale=0.8]{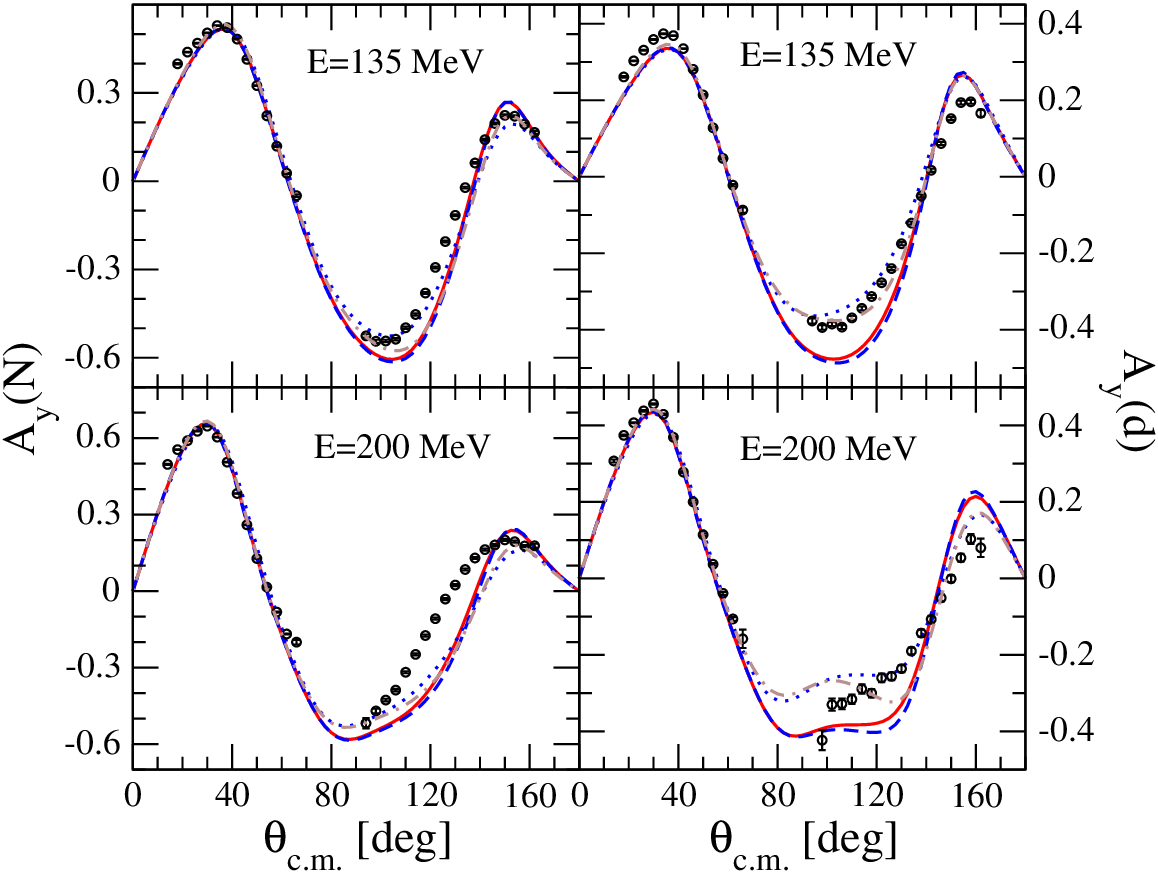}
\caption{
(color online)
The vector analyzing powers  $A_{y}(N)$ and $A_{y}(d)$ 
in elastic nd scattering at the incoming neutron lab. 
energy $E=135$ and $E=200$~MeV.
For description of lines see Fig.\ref{fig1}. 
The pd  data (open circles) are from \cite{indiana}.
}
\label{fig5}
\end{figure}

\begin{figure}
\includegraphics[scale=0.8]{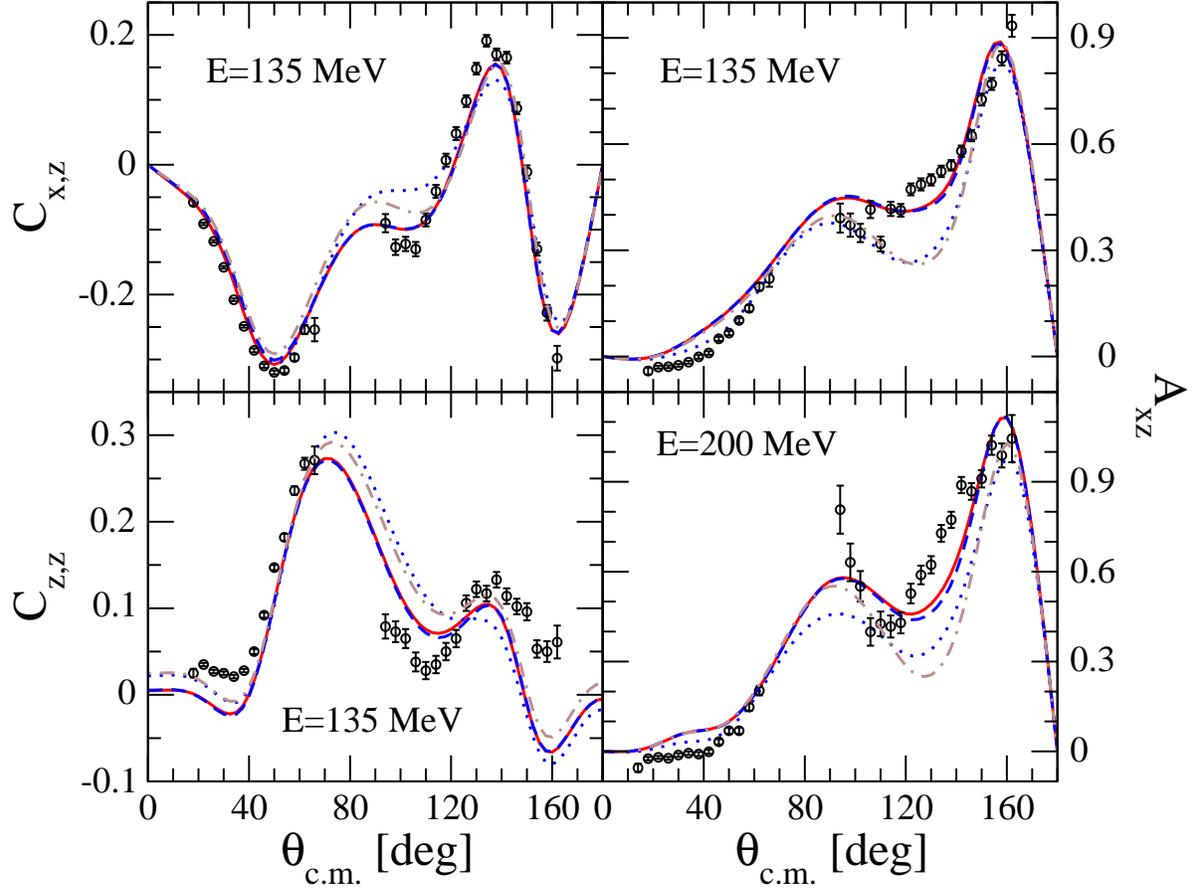}
\caption{
(color online)
The spin correlation coefficients 
$C_{x,z}$ and  $C_{z,z}$  and tensor analyzing power  $A_{xz}$  
in elastic nd scattering at the incoming neutron lab. 
energy $E=135$ and $200$~MeV.  
For description of lines see Fig.\ref{fig1}. 
The pd  data (open circles) are from \cite{indiana}.
}
\label{fig6}
\end{figure}

\begin{figure}
\includegraphics[scale=0.8]{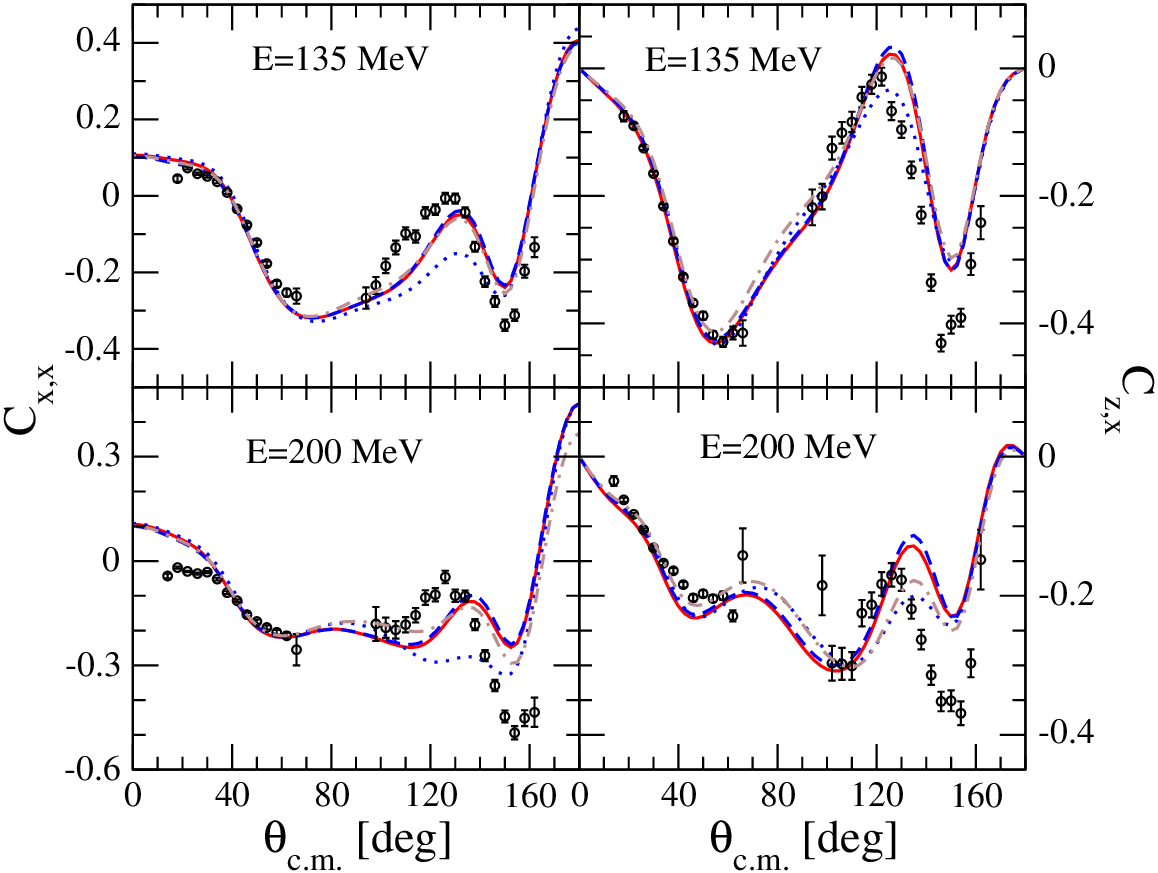}
\caption{
(color online)
The spin correlation coefficients 
$C_{x,x}$ and  $C_{z,x}$  
in elastic nd scattering at the incoming neutron lab. 
energy $E=135$ and $200$~MeV.   
For description of lines see Fig.\ref{fig1}. 
The pd  data (open circles) are from \cite{indiana}.
}
\label{fig7}
\end{figure}

\begin{figure}
\includegraphics[scale=0.8]{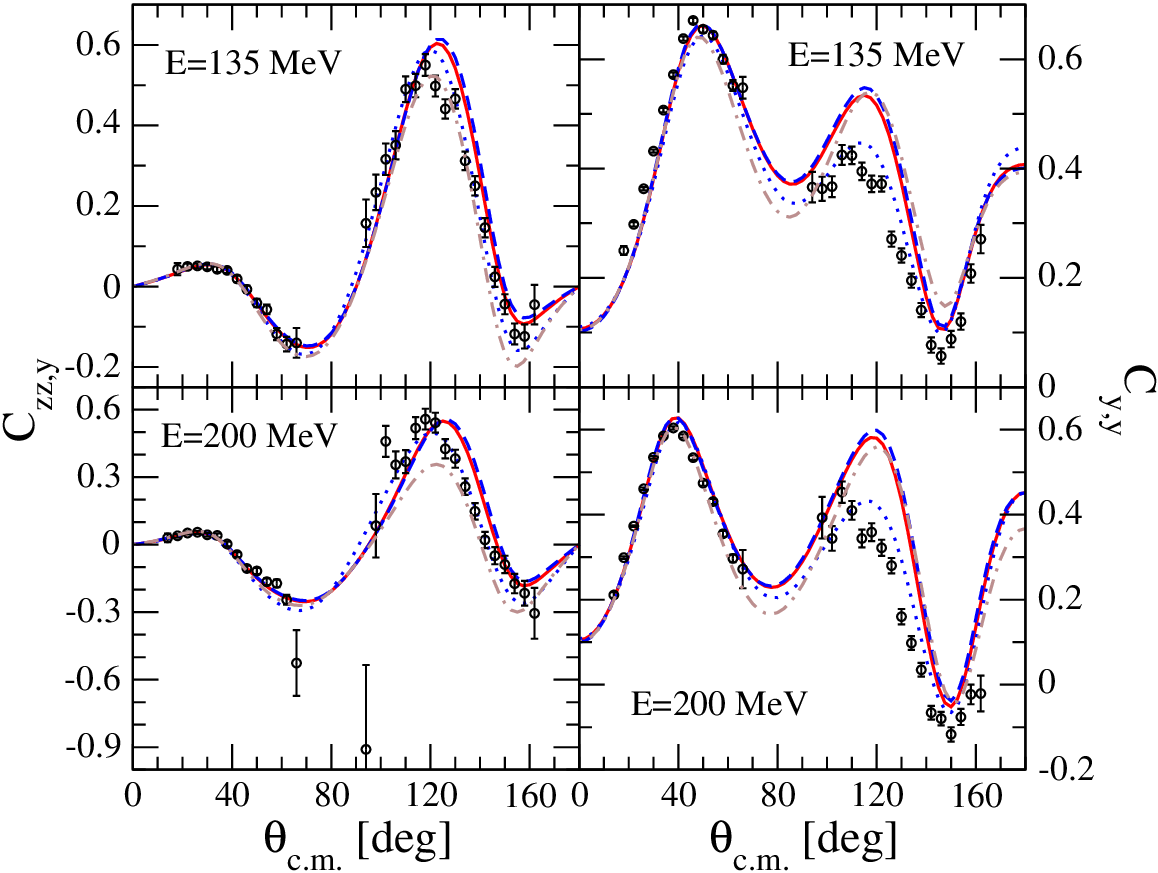}
\caption{
(color online)
The spin correlation coefficients 
$C_{zz,y}$ and  $C_{y,y}$  
in elastic nd scattering at the incoming neutron lab. 
energy $E=135$ and $200$~MeV.  
For description of lines see Fig.\ref{fig1}. 
The pd  data (open circles) are from \cite{indiana}.
}
\label{fig8}
\end{figure}

\begin{figure}
\includegraphics[scale=0.8]{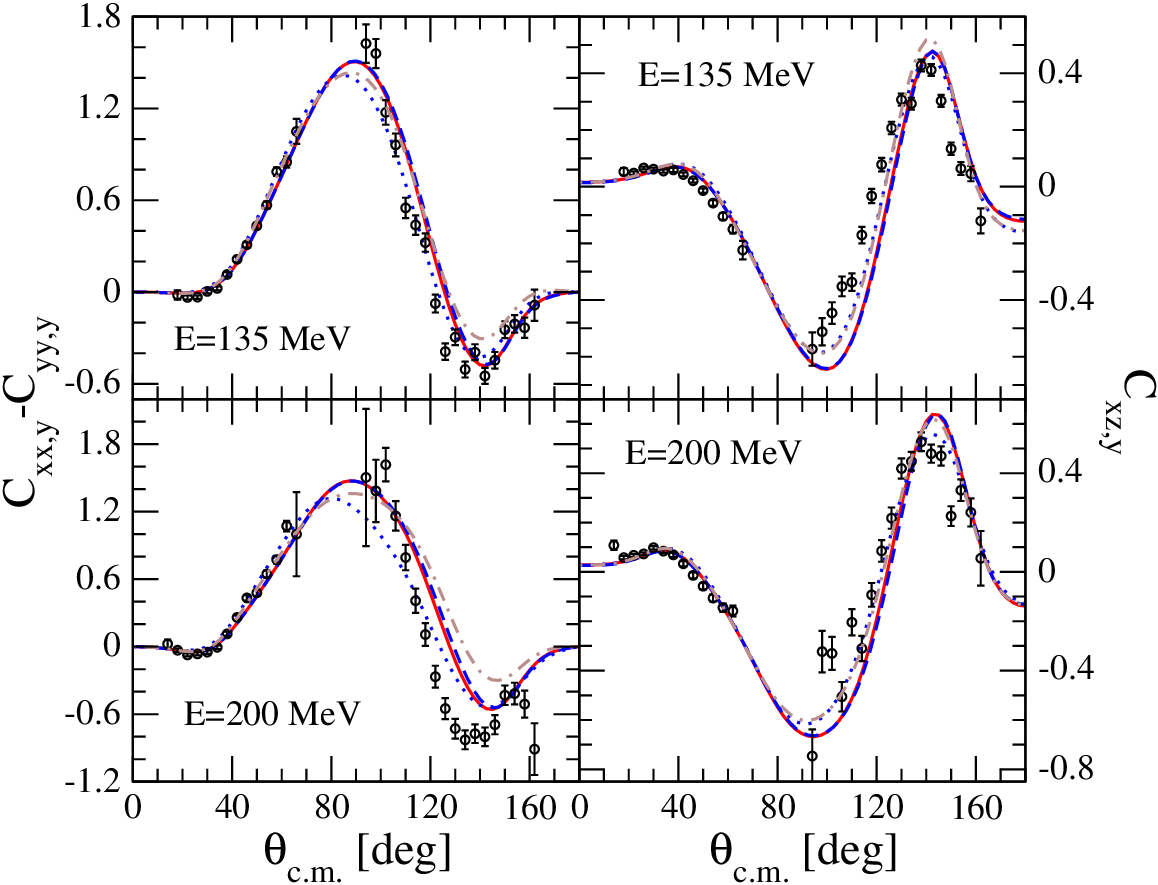}
\caption{
(color online)
The spin correlation coefficients 
$C_{xx,y} - C_{yy,y}$ and  $C_{xz,y}$  
in elastic nd scattering at the incoming neutron lab. 
energy $E=135$ and $200$~MeV.   
For description of lines see Fig.\ref{fig1}. 
The pd  data (open circles) are from \cite{indiana}.
}
\label{fig9}
\end{figure}

\begin{figure}
\includegraphics[scale=0.8]{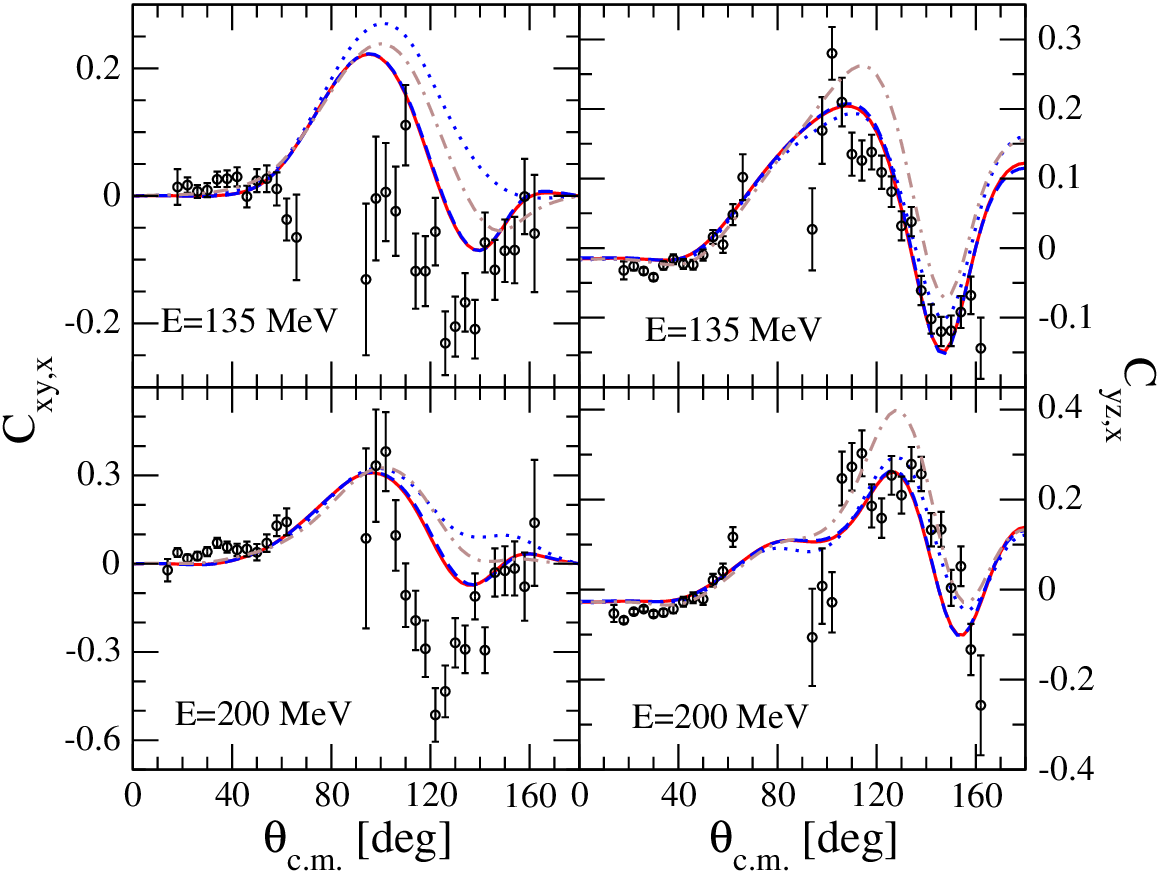}
\caption{
(color online)
The spin correlation coefficients 
$C_{xy,x}$ and  $C_{yz,x}$  
in elastic nd scattering at the incoming neutron lab. 
energy $E=135$ and $200$~MeV.   
For description of lines see Fig.\ref{fig1}. 
The pd  data (open circles) are from \cite{indiana}.
}
\label{fig10}
\end{figure}

\begin{figure}
\includegraphics[scale=0.8]{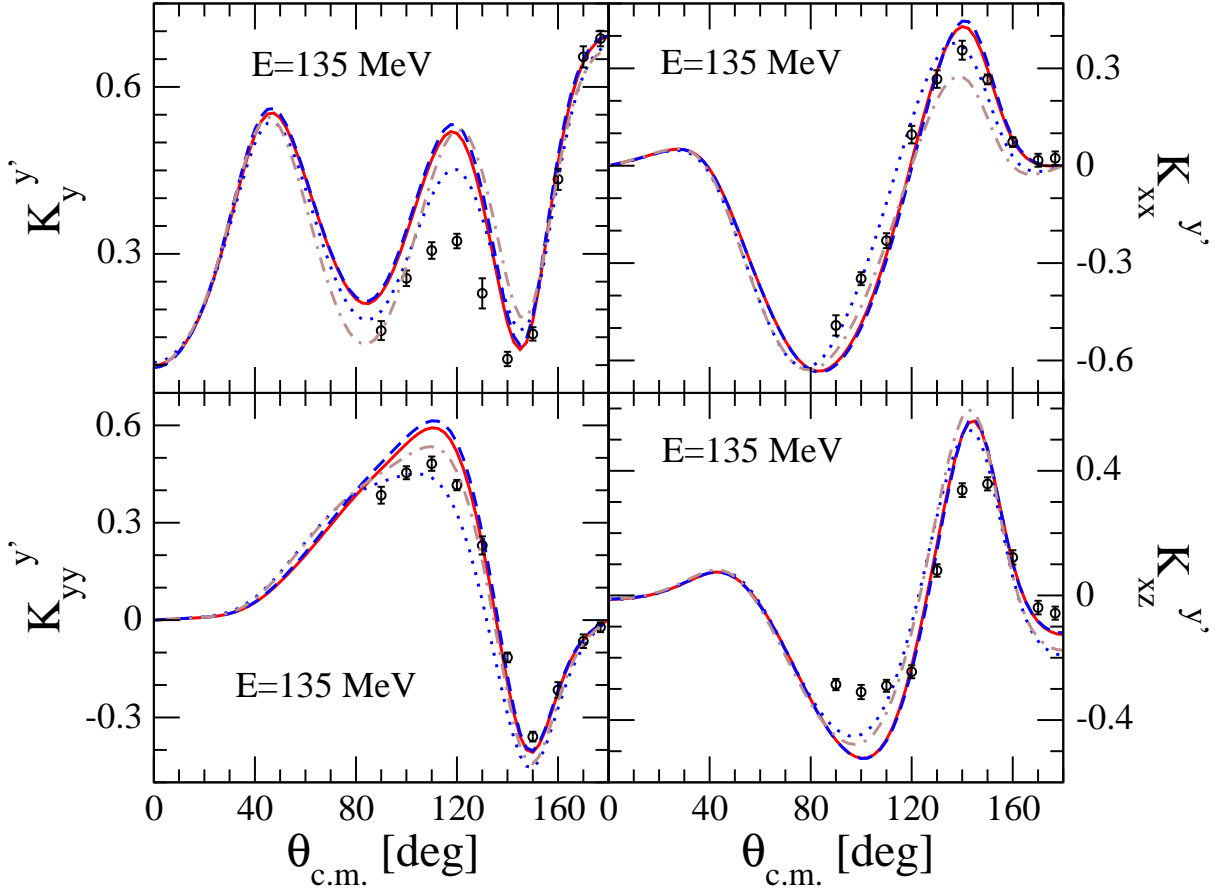}
\caption{
(color online)
The deuteron to neutron polarization 
transfer coefficients  $K_{y}^{y'}$, $K_{xx}^{y'}$, $K_{yy}^{y'}$, and
$K_{xz}^{y'}$ 
in elastic nd scattering at the incoming neutron lab.  
energy $E=135$~MeV. 
For description of lines see Fig.\ref{fig1}. 
The pd  data (open circles) are from \cite{sek_eltransfer}.
}
\label{fig11}
\end{figure}

\begin{figure}
\includegraphics[scale=0.8]{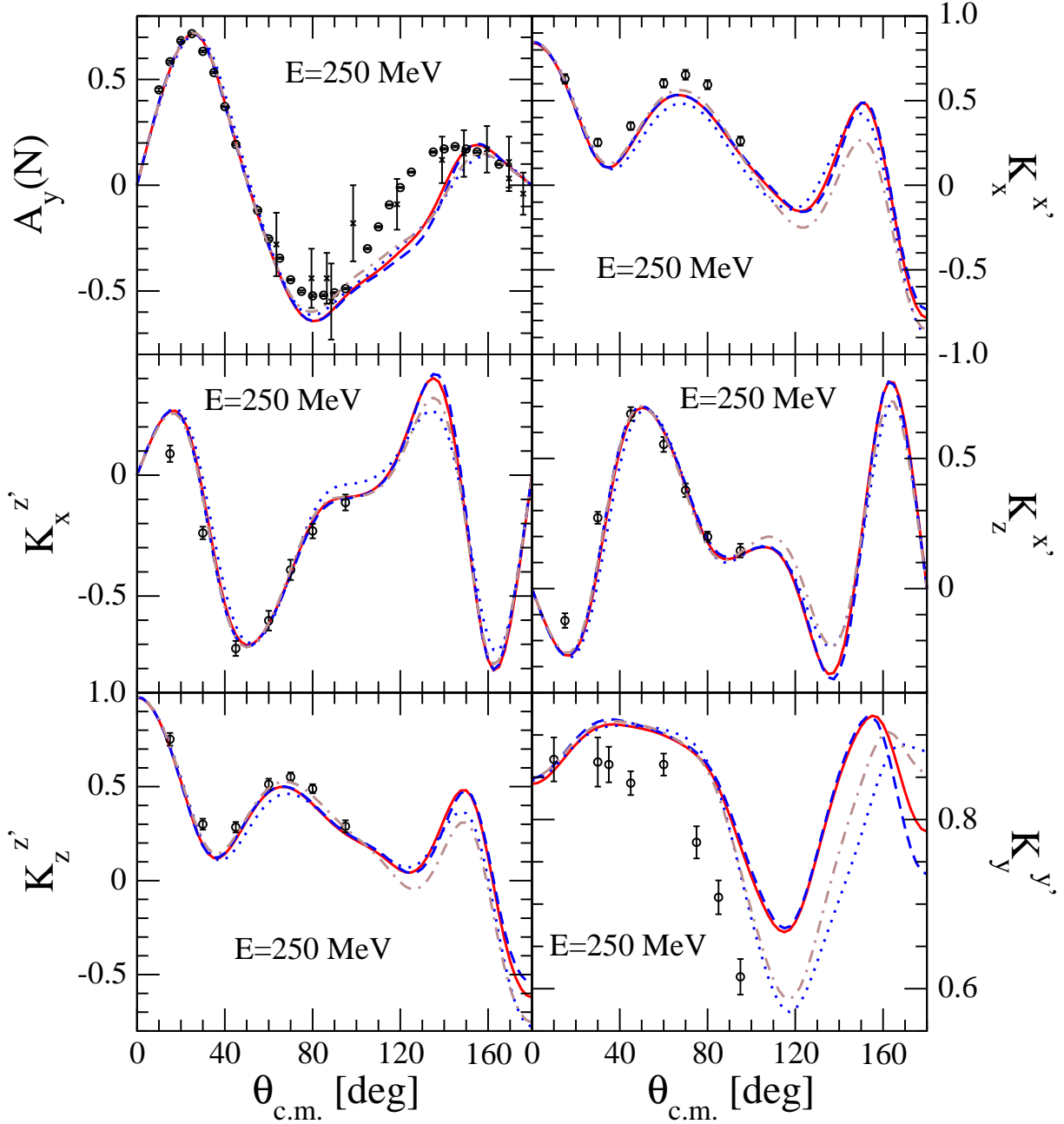}
\caption{
(color online)
The neutron analyzing power $A_y(N)$ and neutron  to neutron  polarization 
transfer coefficients  $K_{x}^{x'}$, $K_{x}^{z'}$, $K_{z}^{x'}$, $K_{z}^{z'}$, and
$K_{y}^{y'}$  
in elastic nd scattering at the incoming neutron lab.  
energy $E=250$~MeV. 
For description of lines see Fig.\ref{fig1}. 
The pd  data (open circles) are from \cite{hat02}. The nd data for
$A_y(N)$ (x-es) are from \cite{maedand}
}
\label{fig12}
\end{figure}

For elastic scattering spin observables effects of relativity, when
only two nucleon forces are acting, were found to be small
\cite{witrel1}.  It is exemplified by nearly overlapping solid (red)
and dashed (blue) lines in Figs.~\ref{fig2}-\ref{fig12}. Adding
three-nucleon force in nonrelativistic calculations leads to
substantial effects for some polarization observables, especially at
higher energies \cite{sek02,wit01}. The resulting picture, however, is
quite complex. Some of those three-nucleon force effects are supported
by the data. For some observables they deteriorate the data
description.

For tensor analyzing powers $A_{xx}$, $A_{yy}$ and $A_{xz}$
relativistic effects are non-negligible even at $70$~MeV (see
Fig.~\ref{fig2}) and clearly increase with increasing energy as seen
in Figs.~\ref{fig3}, \ref{fig4} and \ref{fig6}. When three-nucleon
force is added in the relativistic calculations the resulting effect
depends on the observable and the energy.

For $A_{xz}$ large three-nucleon force effects remain.  At $70$~MeV and
$135$~MeV they are practically identical in magnitude to three-nucleon
force effects found in nonrelativistic calculations and
nonrelativistic and relativistic predictions for $A_{xz}$ at these
energies are practically overlapping (see dotted (blue) and
dashed-dotted (brown) lines in Figs.~\ref{fig2}, \ref{fig3} and
\ref{fig6}). At $200$~MeV,  however, adding three-nucleon force in
relativistic calculations leads to angle dependent modulations of the
magnitude of three-nucleon force effects, similar to that found for
the cross section (see Fig.~\ref{fig6}).

For $A_{xx}$ a drastically different scenario occurs. Large effects of
the TM99 three-nucleon force are seen for that observable in
nonrelativistic calculations at $70$~MeV and $135$~MeV in wide range
of angles and they practically vanish when relativity is included. As a result
the dashed-dotted (brown) line practically overlaps with pure
two-nucleon relativistic and nonrelativistic predictions (see
Fig.~\ref{fig2} and \ref{fig3}).

For $A_{yy}$ (Fig.~\ref{fig2} and \ref{fig3}) the large effects of the
three-nucleon force seen in nonrelativistic calculations are simply
reduced by relativity.  For the tensor analyzing power $A_{zz}$, for
which data exist only at $135$ and $200$~MeV, the influence of
relativity induces both modulation and reduction of nonrelativistic
three-nucleon force effects (Fig.~\ref{fig4}).

The TM99 three-nucleon force acts differently on the nucleon,
$A_y(N)$, and deuteron, $A_y(d)$, vector analyzing powers.  While
three-nucleon force effects for $A_y(N)$ are rather small even at
$250$~MeV (Fig.~\ref{fig5} and \ref{fig12}), for $A_y(d)$ they are
significant (Fig.~\ref{fig2}, \ref{fig3} and \ref{fig5}).  For
$A_y(N)$ and $A_y(d)$, but more clearly displayed due to larger
effects for the deuteron vector analyzing power, both reduction and
modulation of nonrelativistic three-nucleon force effects by
relativity was found.  That reduction and modulation depend on angle
and energy.

A similar picture was found for numerous spin correlation
coefficients, as exemplified by different theoretical predictions
shown in Figs.~\ref{fig6}-\ref{fig10}. Again all scenarios are
available: total reduction by relativity of large three-nucleon force
effects seen in nonrelativistic calculations (e.g. $C_{x,x}$ at $135$
and $200$~MeV for $120^o \le \theta_{c.m.} \le 150^o$ in
Fig.~\ref{fig7}, $C_{y,y}$ at $135$ and $200$~MeV at $120^o \le
\theta_{c.m.} \le 150^o$ in Fig.~\ref{fig8}), practically the same
three-nucleon force effects in nonrelativistic and relativistic
calculations ($C_{z,z}$ at $135$~MeV in Fig.~\ref{fig6}, $C_{xz,y}$ at
$135$ and $200$~MeV in Fig.~\ref{fig9}), angle dependent modulation of
nonrelativistic three-nucleon force effects by relativity ($C_{x,z}$
at $135$~MeV in Fig.~\ref{fig6}, $C_{z,x}$ at $135$ and $200$~MeV in
Fig.~\ref{fig7}, $C_{xy,x}$ and $C_{yz,x}$ at $135$ and $200$~MeV in
Fig.~\ref{fig10}).

The polarization transfer coefficients are not exceptions; also for
them a similar complex influence of relativity on nonrelativistic
three-nucleon force effects have been found as shown in
Figs.~\ref{fig11} and \ref{fig12}.

The comparison of nonrelativistic predictions based on $2\pi$-exchange
three-nucleon force's revealed for spin observables a complex, angle
and energy dependent pattern of discrepancies between data and theory
\cite{wit01,sek02,sek_eltransfer,maedand,hat02}.  The nontrivial
interplay between the $2\pi$-exchange three-nucleon forces and
relativity suggests that the inclusion  of further three-nucleon force 
mechanisms, like forces of shorter range, is needed to improve the
 description  of elastic scattering polarisation  data.

\subsection{Breakup}
\label{breakup}

\begin{figure}
\includegraphics[scale=0.8]{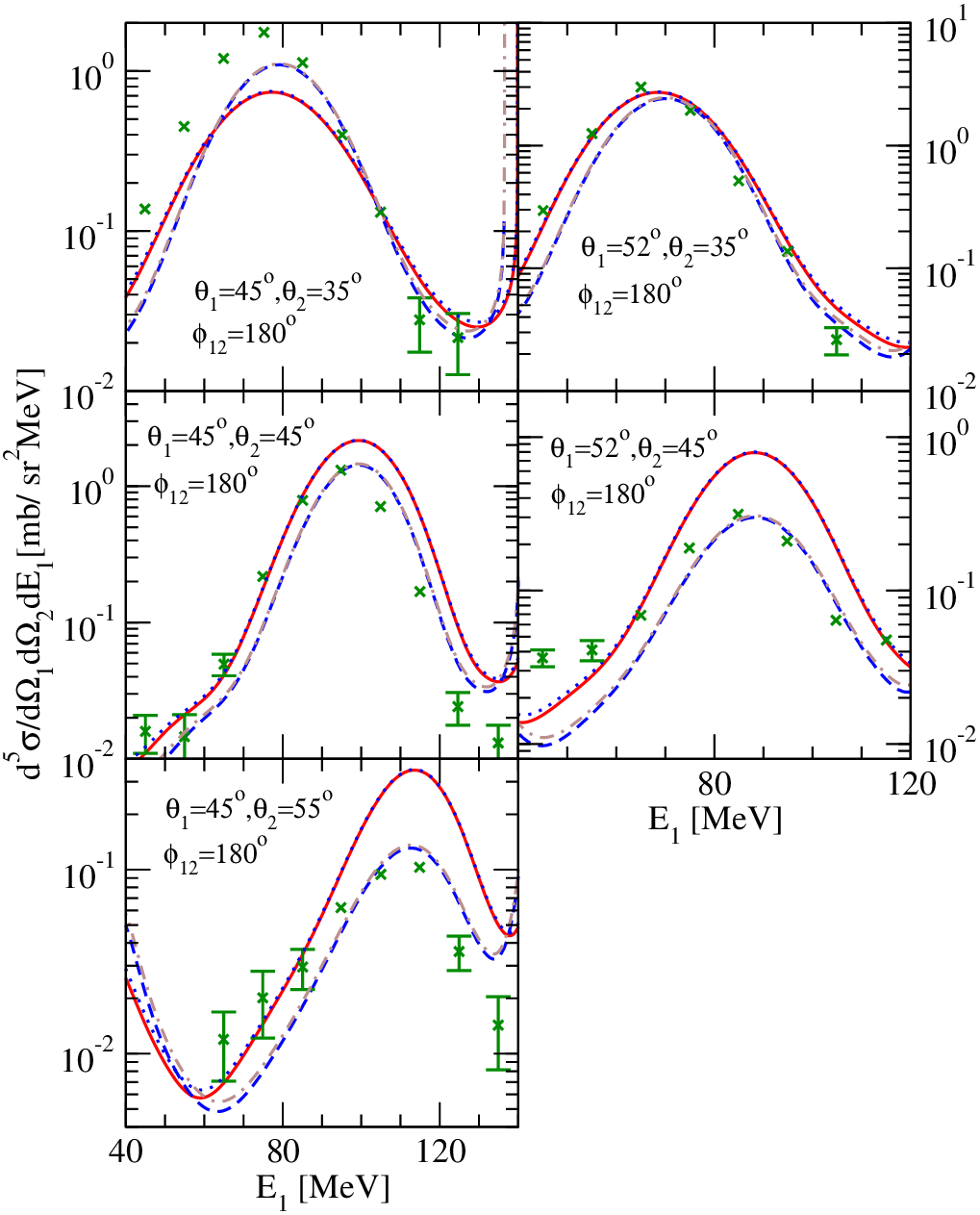}
\caption{
(color online) 
The five-fold cross section
  $d^5\sigma/d\Omega_1d\Omega_2dE_1^{lab}$ for the breakup reaction
  d(n,np)n at $E_n^{lab}=200$~MeV and fixed angles of outgoing nucleons
1 and 2 as indicated in the figures. 
For description of lines see Fig.\ref{fig1}. 
The d(p,pn)p data (x-es)
are from \cite{brdat}.
}
\label{fig13}
\end{figure}

Theoretical study of exclusive breakup reaction performed at different
incoming nucleon energies revealed regions of breakup phase-space
where large three-nucleon force effects have been found
\cite{zoln1}. The effects, similarly to elastic scattering, generally
increase with energy. With increasing energy also the effects of
relativity increase \cite{witrel3,skibbr}, revealing for exclusive
breakup cross section a characteristic pattern when viewed as a
function of the angles of detected nucleons. Largest effects where
found when two of three outgoing nucleons are detected coplanarly on
both sides of the beam. Keeping one of the detectors at a constant
position and changing the polar angle of the second, regions of phase
space were found in which nonrelativistic breakup cross section was
increased or decreased by relativity \cite{skibbr}. In these specific
configurations effects of three-nucleon force's on breakup cross
section, both in norelativistic as well as in relativistic
calculations, are practically negligible (see Fig.~\ref{fig13}).

\begin{figure}
\includegraphics[scale=0.8]{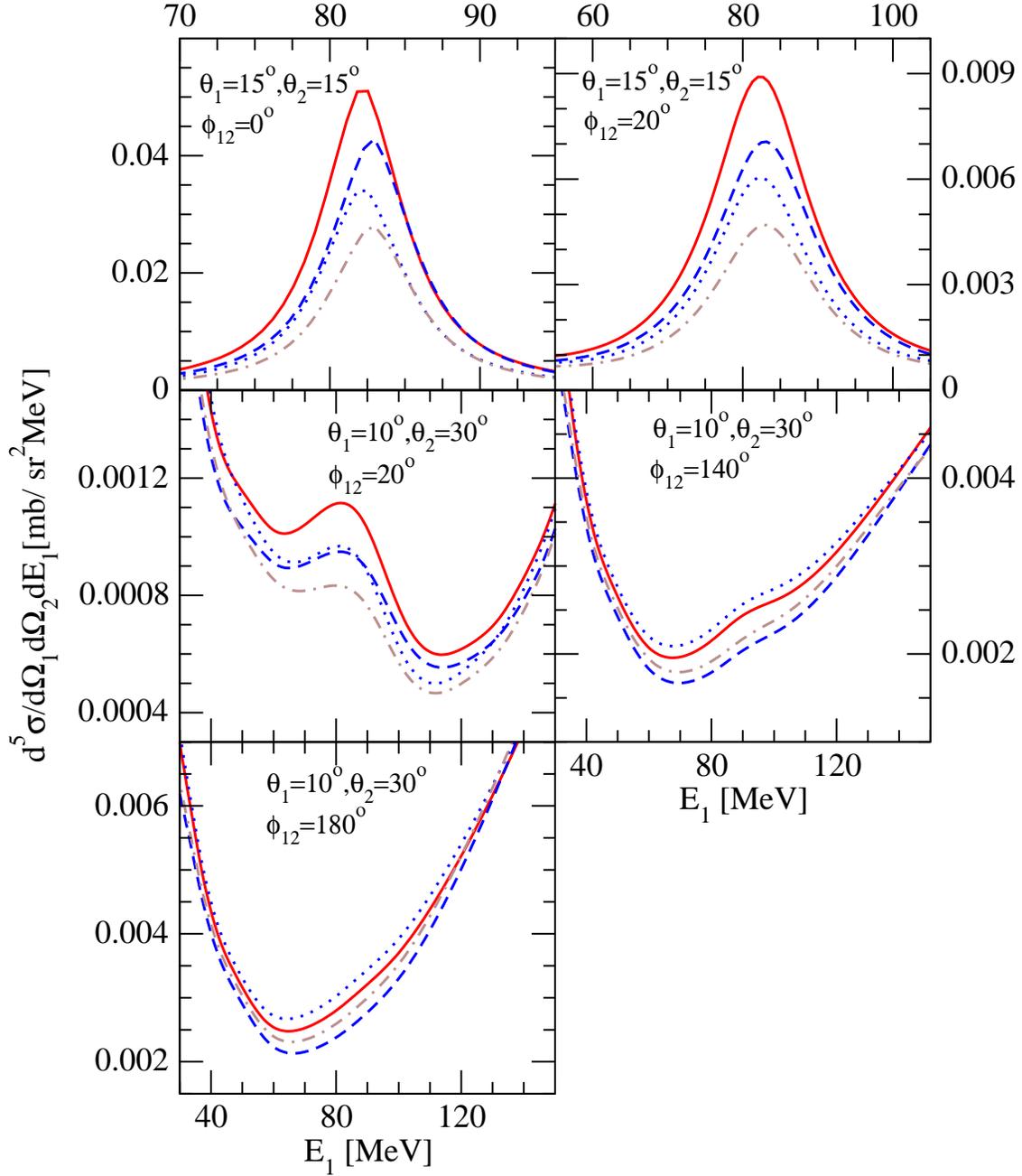}
\caption{
(color online) 
The five-fold cross section
  $d^5\sigma/d\Omega_1d\Omega_2dE_1^{lab}$ for the breakup reaction
  d(n,np)n at $E_n^{lab}=200$~MeV and fixed angles of outgoing nucleons
1 and 2 as indicated in the figures. 
For description of lines see Fig.\ref{fig1}. 
}
\label{fig14}
\end{figure}

Due to richness of the breakup phase-space also geometrical
configurations can be found where both, three-nucleon force and
relativistic effects are significant. Exclusive cross sections in some
of these configurations are shown as a function of the laboratory
energy of one of the outgoing and detected nucleons in
Fig.~\ref{fig14} for neutron-deuteron breakup at $200$~MeV. It is seen
that including relativity reduces slightly the magnitude of
three-nucleon force effects observed in nonrelativistic calculations.

\begin{figure}
\includegraphics[scale=0.8]{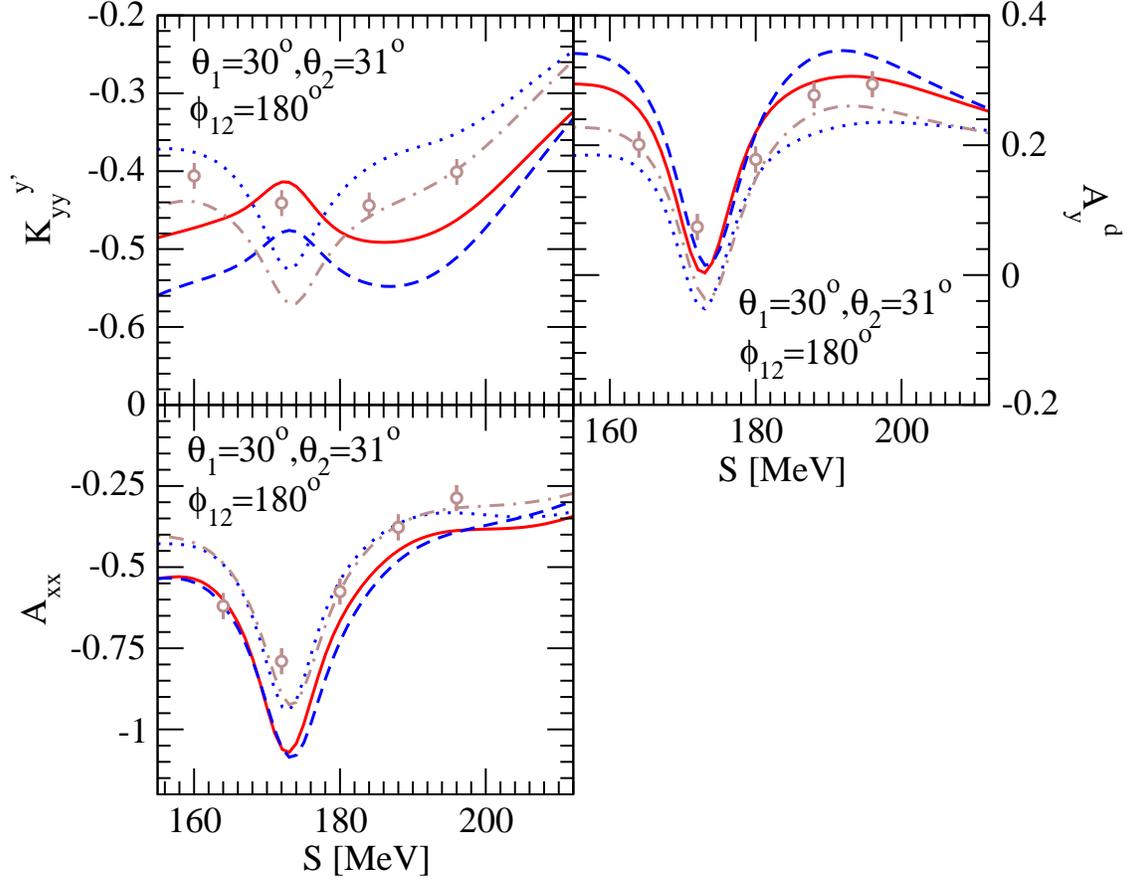}
\caption{
(color online) 
The polarization transfer coefficient $K_{yy}^{y'}$ and deuteron
analyzing powers $A_y^d$ and $A_{xx}$ in  the breakup reaction
  n(d,nn)p at incoming deuteron lab. energy $E_d^{lab}=270$~MeV, shown 
as a function of the  S-curve arc length.  
For description of lines see Fig.\ref{fig1}. 
The $270$~MeV dp data (circles) are from ref.~\cite{brkim}.
}
\label{fig15}
\end{figure}

Relativity changes also the magnitude of three-nucleon force effects
seen in nonrelativistic calculations for breakup polarization
observables.  We exemplify that in Fig.~\ref{fig15} at three
configurations of exclusive dp breakup at $E_d^{lab}=270$~MeV, for
which data have been taken \cite{brkim}. Again, influence of
relativity on magnitude of three-nucleon force effects change with
configuration as shown in Fig.~\ref{fig15} along  the S-curve arc
length. Especially interesting is the case of polarization-transfer
coefficient from the deuteron to the nucleon, $K_{yy}^{y'}$, for which
inclusion of TM99 three-nucleon force changes completely the
S-dependence found in case when only two nucleon-forces were
acting. The effect of three-nucleon force is further modified slightly
by relativity resulting in a better reproduction of data.

\section{Summary and outlook}
\label{summary}

We extended our relativistic formulation of three-nucleon Faddeev
equations to include also three-nucleon force.  The relativistic
features are the relativistic form of the free propagator, the change
of the nucleon-nucleon potential caused by the boost of the two
nucleon subsystem, and the modification of the permutation operators.
In present study we neglected Wigner spin rotations induced by these
boosts since investigations based on two-nucleon forces only have
shown that their effects are negligible.  For the momentum-space basis
we used the relative momentum of two free nucleons in their
c.m. system together with their total momentum in the three nucleon
c.m. system, which in this frame is the negative momentum of the
spectator nucleon.  Such a choice of momenta is adequate for
relativistic kinematics and allows to generalize the nonrelativistic
approach used to solve the nonrelativistic three nucleon Faddeev
equation to the relativistic case in a more or less straightforward
manner.  That relative momentum in the two-nucleon subsystem is a
generalization of the standard nonrelativistic Jacobi momentum
$\mathbf{p}$.  We numerically solved these equations for
neutron-deuteron scattering including relativistic features and/or
three-nucleon force at the neutron lab energies $E_n^{lab} = 70$,
$135$, $200$ and $250$~MeV.  As dynamical input we took the
nonrelativistic nucleon-nucleon potential CD~Bonn and generated in the
two nucleon center of momentum system an exactly on-shell equivalent
relativistic interaction.  As a three-nucleon force we took the
$2\pi$-exchange TM99 force.

By comparing our relativistic calculations without and with
the three-nucleon force included we studied influence of relativity on
three-nucleon force effects. In studies with two-nucleon forces only
it was found that significant relativistic effects for the elastic
scattering cross section appear at higher energies and they are
restricted only to the very backward angles where relativity increases
the nonrelativistic cross section. At other angles the effects are
small.  Also for spin observables, analyzing powers, spin correlation
coefficients and spin transfer coefficients, no significant changes
due to relativity have been found when only two-nucleon forces were
acting.  The similar picture was found for breakup, however, in that
case significantly larger effects for the cross section in specific
regions of the breakup phase-space have been found.

The results obtained in the present study document that this picture
changes dramatically when in addition to the two-nucleon force in
a relativistic treatment also a three-nucleon force is acting.  For the
elastic scattering large changes of the cross section at higher
energies, caused by three-nucleon force in large region of angles
ranging from around minimum of the cross section up to very backward
angles, are further significantly modulated by relativity. Also such
modulation in a large, similar to that for the cross section, range of
angles have been found for numerous polarization observables. In that
case every conceivable scenario of modulations was observed: from
wiping out large three-nucleon force effects found in nonrelativistic
calculations to their modulations with energy and angle, with strong
amplification or reduction of their magnitude.  Also for exclusive
breakup cross section and polarisation breakup observables in some
geometries the relativity influences effects induced by three-nucleon
forces. Thus also for that reaction the relativistic treatment when
three-nucleon forces are acting is required for proper interpretation
of data.

The comparison of our nonrelativistic theory with existing elastic
scattering cross section and polarisation data exhibits at the higher
energies clear discrepancies.  The discrepancies between the theory
based on pairwise forces only and data are largest in the region
starting from the cross section minimum around $\theta_{c.m.} \approx
130^{\circ}$ up to $\theta_{c.m.} \approx 180^{\circ}$.  At energies
up to about $\approx 135$~MeV these discrepancies can be removed when
current three-nucleon forces, mostly of
$2\pi$-exchange character~\cite{TM,uIX}, are included in the nuclear
Hamiltonian.  At the higher energies, however, a significant part of
the discrepancy remains and increases further with increasing energy.
Especially complex picture exists for spin observables. Here adding
$2\pi$-exchange three-nucleon force into nonrelativistic calculations
leads to effects which depend on observable. They can be large or
negligible, change their magnitude with energy and angle. Similarly
to the elastic scattering cross section even after inclusion of
three-nucleon force some of the discrepancies remain and increase with
increasing energy.  This indicates that additional three nucleon
forces should be added to the $ 2\pi$-exchange type forces.  Natural
candidates in the traditional meson-exchange picture are exchanges
like $\pi-\rho$ and $\rho-\rho$.  This has to be expected since in
$\chi$PT \cite{epel2002} in the order in which nonvanishing
three-nucleon force's appear the first time there are three topologies
of forces, the $2\pi$-exchange, a one-pion exchange between one
nucleon and a two-nucleon contact interaction and a pure three nucleon
contact interaction. They are of the same order and have to be kept
together. Therefore it appears very worthwhile to pursue a strategy
adding in the traditional meson exchange picture further three nucleon
forces.  Results presented here show that relativistic effects based
on relativistic kinematics and boost effects of the nucleon-nucleon
force play an important role in building up the magnitude of
three-nucleon force effects.  That gives hope, that taking the proper
three-nucleon force into relativistic Faddeev calculations one will be
able to improve the description of higher energy data for cross
section and polarization observables.

\section*{Acknowledgments}
This work has been supported by the Polish 2008-2011 science funds as
the research project No. N N202 077435, by the Helmholtz
Association through funds provided to the virtual institute ``Spin and
strong QCD''(VH-VI-231),  and by
  the European Community-Research Infrastructure
Integrating Activity
``Study of Strongly Interacting Matter'' (acronym HadronPhysics2,
Grant Agreement n. 227431)
under the Seventh Framework Programme of EU.   
 W. P. is supported by the U.S. Department of Energy, contract 
DE-FG02-86ER40286. 
H. W. would like to thank the Kyushu
University for hospitality and support during his stay in this
institution.  The numerical calculations were performed on the
supercomputer cluster of the JSC, J\"ulich, Germany.

\appendix
\section{Direct recalculation of the partial-wave projected 
three-nucleon force matrix elements from ($p,q$)- to ($k,q$)- based basis}
\label{app_a}

We start from the matrix element of a three-nucleon force
$\left\langle p,q, \alpha | V_4^{(1)} (1 + P) | p',q', \alpha'
\right\rangle$ in a partial wave basis used in norelativistic
calculations with standard Jacobi momenta (p,q) \cite{book} and would
like to get the matrix element $\left\langle k,q,\alpha | V_4^{(1)} (1
+ P) | k',q',\alpha'\right\rangle$ with $p$ and $p'$ replaced by the
relative momenta of nucleons $2$ and $3$, $k$ and $k'$, in their
two-nucleon center of momentum system \cite{witrel1,witrel2}.

Using completeness of partial wave states one has
\begin{eqnarray}
  \left\langle k,q,\alpha  | V_4^{(1)} (1 + P) | k',q', \alpha' 
\right\rangle
  &=& \sum\limits_{\tilde \alpha } \int  d\tilde p\tilde p^2
        \int d\tilde q\tilde q^2
 \sum\limits_{\tilde \alpha'} \int d\tilde p' \tilde p'^2  \int
 d\tilde q' \tilde q'^2 \left\langle k,q,\alpha 
  |  \tilde p, \tilde q ,\tilde \alpha  \right\rangle    \cr
 && \left\langle \tilde p, \tilde q, \tilde \alpha  | V_4^{(1)} (1 + P) |
 \tilde p', \tilde q', \tilde \alpha' \right\rangle \left\langle
 \tilde p', \tilde q', \tilde \alpha'
  |  k',q', \alpha' \right\rangle   ~.
\label{a1}
\end{eqnarray}
The partial wave state used in relativistic calculations 
 $|\mathbf{P},  k,q,\alpha  \rangle$ corresponding to the total three-nucleon
center of momentum, $\mathbf{P}=\mathbf{0}$, is given by \cite{witrel1}
\begin{eqnarray}
  |\mathbf{P},  k,q,\alpha  \rangle  &=&  
| \mathbf{P}, k,q(l,s)j(\lambda \frac {1}{2})I(jI)JM;
(t\frac {1}{2})T \rangle  = \sum\limits_{\mu_1 \mu_2 \mu_3 }
         \sum\limits_{\mu_2 '\mu_3 '} 
\sum\limits_{\mu_s \mu_l \mu_{\lambda}  \mu_I }  \int d\hat \mathbf{q} \int 
d\hat \mathbf{k}  Y_{l\mu_l } 
(\hat \mathbf{k}) N(\mathbf{q}_2 ,\mathbf{q}_3 )  \cr 
&&  (\frac {1}{2} \frac {1}{2},s | \mu_2, \mu_3, \mu_s ) (l,s,j |
\mu_l, \mu_s, \mu )\cr
&& D_{\mu_2 '\mu_2 }^{1/2} (R_{wc}(B_c(-q),{k}_2(\mathbf{q}_2,\mathbf{q}_3 ))
D_{\mu_3 '\mu_3 }^{1/2} (R_{wc}(B_c(-q), {k}_3(\mathbf{q}_2,\mathbf{q}_3 )) \cr
&&  Y_{\lambda \mu_\lambda  } (\hat \mathbf{q})(\lambda ,\frac {1}{2}, I | 
\mu_{\lambda},  \mu_1, \mu_I ) (j,I,J | \mu, \mu_I, M) \cr
&& | \mathbf{q} + \frac{1}{3} \mathbf{P},  \mu_1 \rangle 
|  \mathbf{q}_2 (\mathbf{k}, - \mathbf{q}) + \frac{1}{3} \mathbf{P}, \mu_2 ' 
\rangle | \mathbf{q}_3 ( - \mathbf{k}, - \mathbf{q}) + \frac{1}{3} \mathbf{P},
 \mu_3 ' \rangle | (t,\frac {1}{2}),T \rangle 
\label{a2}
\end{eqnarray}
where 
\begin{eqnarray}
 N^2 ({\mathbf{q}}_2, {\mathbf{q}}_3) &\equiv& 
\vert~ { {\partial(\mathbf{q}_2~\mathbf{q}_3)} 
\over { \partial(\mathbf{P}_{NN}~\mathbf{k}) } } ~ \vert = 
 {{\bar M}_0 \over{\omega_{{\bar M}_0}({P}_{NN}) }} ~ 
 {\omega_{q_2} \over{\omega_{k} }} ~
 {\omega_{q_3} \over{\omega_{k} }}  
\label{a2_a}
\end{eqnarray}
is the Jacobian for the Lorentz transformation from 
$({\mathbf{q}}_2, {\mathbf{q}}_3)$ to  $(\mathbf{P}_{NN}, \mathbf{ k})
= (-\mathbf{q}, \mathbf{k})$,  
$\omega_k = \sqrt{m^2 + k^2}$, 
${\bar M}_0=2\omega_k =\omega_{q_2} + \omega_{q_3}$, and 
$\omega_{{\bar M}_0}(P_{NN}) = \sqrt{{\bar M}_0^2 + P_{NN}^2}$. 
The momentum $\mathbf{k}_2(\mathbf{q}_2,\mathbf{q}_3) = \mathbf{k}$
and  $\mathbf{k}_3(\mathbf{q}_2,\mathbf{q}_3)= 
-\mathbf{k}_2(\mathbf{q}_2,\mathbf{q}_3)$.

The nonrelativistic partial-wave state 
$| \mathbf{P}', \tilde p, \tilde q, \tilde \alpha  \rangle$ 
 with standard Jacobi momenta is given by
\begin{eqnarray}
  |\mathbf{P}', \tilde p, \tilde q, \tilde \alpha  \rangle  &=&
\sum\limits_{\tilde \nu_1 ,\tilde \nu_2, \tilde \nu_3 }
           \sum\limits_{\tilde \nu_s, \tilde \nu_l, \tilde \nu_{\lambda}, \tilde \nu_I }
  \int d\hat {\tilde \mathbf{p}} \int d\hat {\tilde \mathbf{q}} 
Y_{\tilde l \tilde \nu_l } 
(\hat {\tilde \mathbf{p}}) Y_{\tilde \lambda \tilde \nu_{\lambda} } 
(\hat {\tilde \mathbf{q}}) \cr
  &&(\frac {1}{2}, \frac{1}{2}, \tilde s | \tilde \nu_2, 
\tilde \nu_3, \tilde \nu_s)
(\tilde l, \tilde s, \tilde j | \tilde \nu_l, \tilde \nu_s, \tilde \nu) 
(\tilde \lambda, \frac {1}{2}, \tilde I | \tilde \nu_{\lambda}, 
 \tilde \nu_1, \tilde \nu_I )
(\tilde j ,\tilde I, \tilde J | \tilde \nu, \tilde \nu_I, \tilde M)
 \cr
&&  |  {\tilde \mathbf{q}} + \frac{1}{3} \mathbf{P}', \tilde \nu_1 \rangle 
| {\tilde \mathbf{q}}_2^{nr}  + \frac{1}{3} \mathbf{P}', \tilde \nu_2  \rangle 
| {\tilde \mathbf{q}}_3^{nr}  + \frac{1}{3} \mathbf{P}', 
\tilde \nu_3  \rangle | 
 (\tilde t \frac{1}{2}) \tilde T \rangle  
\label{a3}
\end{eqnarray}
where in the  three-nucleon center of momentum system the nonrelativistic 
momenta $ {\tilde \mathbf{q}}_2^{~nr}$ 
and $ {\tilde  \mathbf{q}}_3^{~nr}$ of the nucleons 2 and 3 
 are given by standard Jacobi momenta $ {\tilde \mathbf{p}}$ 
and $ {\tilde \mathbf{q}}$ as
\begin{eqnarray}
 {\tilde \mathbf{q}}_2^{~nr} &=& {\tilde \mathbf{p}} 
- \frac{ {\tilde \mathbf{q}}}{2} \cr 
 {\tilde \mathbf{q}}_3^{~nr} &=& - {\tilde \mathbf{p}} 
- \frac{ {\tilde \mathbf{q}}}{2} ~.
\label{a3_a}
\end{eqnarray}
That leads to the scalar product $\left\langle {{\mathbf{P}, k,q,\alpha }} | 
 \mathbf{P}', {{\tilde p,\tilde q,\tilde \alpha }} \right\rangle$
\begin{eqnarray}
  \langle \mathbf{P}, k,q, \alpha  &|& \mathbf{P}',
 \tilde p, \tilde q, \tilde \alpha  \rangle  =
 \sum\limits_{\mu _1 \mu _2 \mu _3}
 \sum\limits_{\mu _2 '\mu _3 '} \sum\limits_{\mu_s \mu_l \mu_\lambda  \mu_I
 } \int d\hat \mathbf{q}
\int d\hat \mathbf{k}  Y_{l \mu_l}^* (\hat \mathbf{k})
N(\mathbf{q}_2 ,\mathbf{q}_3)  
\cr  
&& (\frac {1}{2}, \frac {1}{2}, s | \mu_2, \mu_3, \mu_s )
 (l,s,j | \mu_l, \mu_s, \mu )  \cr
 && D_{\mu_2 '\mu_2 }^{1/2*} (R_{wc}(B_c(-q),{k}_2( \mathbf{q}_2,\mathbf{q}_3 ))
    D_{\mu_3 '\mu_3 }^{1/2*} (R_{wc}(B_c(-q), {k}_3(
    \mathbf{q}_2,\mathbf{q}_3 ))  \cr
&& Y_{\lambda \mu_\lambda}^* (\hat \mathbf{q})(\lambda, \frac {1} {2}, I | 
 \mu_{\lambda},  \mu_1, \mu_I ) (j,I,J | \mu, \mu _I, M) \cr
&&  \sum\limits_{\tilde \nu_s \tilde \nu_l \tilde \nu_{\lambda} \tilde \nu_I } \int
  d\hat {\tilde \mathbf{p}} 
\int d\hat {\tilde \mathbf{q}} Y_{\tilde l \tilde \nu_l} 
(\hat {\tilde \mathbf{p}})
Y_{\tilde {\lambda}, \tilde {\nu}_{\lambda}} (\hat {\tilde \mathbf{q}}) 
(\frac {1} {2}, \frac {1}{2}, \tilde s | \mu_2', \mu_3', \tilde {\nu}_s )
(\tilde l, \tilde s, \tilde j | \tilde \nu_l, \tilde \nu_s, \tilde \nu) \cr 
&& (\tilde \lambda, \frac {1}{2}, \tilde I 
| \tilde \nu_{\lambda},  \mu_1, \tilde \nu_I)
(\tilde j, \tilde I \tilde J | \tilde \nu, \tilde \nu_I, \tilde M) \cr
&&  \delta (\mathbf{q} -  {\tilde \mathbf{q}}  
+ \frac{1}{3}(\mathbf{P} - \mathbf{P}')  ) 
\delta (\mathbf{q}_2 
(\mathbf{k}, - \mathbf{q}) - 
 {\tilde \mathbf{q}}_2^{~nr} + \frac{1}{3}(\mathbf{P} - \mathbf{P}') ) \cr
&&\delta (\mathbf{q}_3 ( - \mathbf{k}, - \mathbf{q}) 
- \mathbf{ {\tilde q}}_3^{~nr} 
+ \frac{1}{3}(\mathbf{P} - \mathbf{P}') )  
  \left\langle (t \frac{1}{2})T  | (\tilde t \frac{1}{2}) \tilde T 
 \right\rangle  ~.
\label{a4}
\end{eqnarray}
That matrix element should be proportional to $\delta _{J\tilde J} 
\delta _{M\tilde M}$ and independent from $M$. Thus
\begin{eqnarray}
  \langle \mathbf{P}, k,q,\alpha  &|& \mathbf{P}', 
  \tilde p, \tilde q, \tilde \alpha  \rangle  =
 \frac{1} {2J+1} \sum\limits_{M}
 \sum\limits_{\mu_1, \mu_2, \mu_3 } \sum\limits_{\mu_2 '\mu_3 '} 
\sum\limits_{\mu_s, \mu_l, \mu_{\lambda},  \mu_I} 
\int d\hat \mathbf{q}\int d\hat \mathbf{k} 
Y_{l\mu_l}^* (\hat \mathbf{k}) N(\mathbf{q}_2 ,\mathbf{q}_3 )\cr  
&& (\frac {1}{2}, \frac {1}{2}, s | \mu_2, \mu_3, \mu_s ) (l,s,j | 
\mu_l, \mu_s, \mu )  \cr
 && D_{\mu_2 '\mu_2}^{1/2*} (R_{wc}(B_c ( -q),{k}_2( \mathbf{q}_2,\mathbf{q}_3 ))
D_{\mu_3 '\mu_3 }^{1/2*} 
(R_{wc}(B_c ( -q), {k}_3( \mathbf{q}_2,\mathbf{q}_3 ))  \cr
&&  Y_{\lambda \mu_{\lambda}}^* (\hat \mathbf{q})(\lambda, \frac{1}{2}, I | 
\mu_{\lambda},  \mu_1, \mu_I ) (j,I,J | \mu,  \mu_I, M) \cr
&&  \sum\limits_{\tilde \nu _s \tilde \nu_l \tilde \nu_{\lambda} \tilde \nu_I } \int
  d\hat {\tilde \mathbf{p}} 
\int d\hat {\tilde \mathbf{q}} Y_{\tilde l \tilde \nu_l} 
(\hat {\tilde \mathbf{p}})
Y_{\tilde \lambda \tilde \nu_{\lambda}} (\hat {\tilde \mathbf{q}}) 
(\frac {1}{2}, \frac {1}{2}, \tilde s | \mu_2', \mu_3', \tilde {\nu}_s )
(\tilde l, \tilde s, \tilde j | \tilde \nu_l, \tilde \nu_s, \tilde \nu) \cr 
&& (\tilde \lambda, \frac {1}{2}, \tilde I | \tilde \nu_{\lambda},  \mu_1, 
\tilde \nu_I ) 
(\tilde j, \tilde I, J | \tilde \nu, \tilde \nu_I, M)  \cr
&&  \delta (\mathbf{q} -  {\tilde \mathbf{q}} 
+ \frac{1}{3}(\mathbf{P} - \mathbf{P}') )
\delta (\mathbf{q}_2 
(\mathbf{k}, - \mathbf{q}) - 
 {\tilde \mathbf{q}}_2^{~nr} + \frac{1}{3}(\mathbf{P} - \mathbf{P}') )\cr
&&\delta(\mathbf{q}_3 
( - \mathbf{k}, - \mathbf{q}) -  {\tilde \mathbf{q}}_3^{~nr} 
+ \frac{1}{3}(\mathbf{P} - \mathbf{P}') )  
  \langle (t \frac {1}{2}T)| (\tilde t \frac{1}{2}) \tilde T \rangle  ~.
\label{a5}
\end{eqnarray}

The momenta of nucleons 2 and 3 in the three-nucleon center 
of momentum system are given
through 
their two-nucleon center of momentum relative 
momentum  $\mathbf{k}$ and the momentum of the
spectator nucleon 1, $\mathbf{q}$, by
\begin{eqnarray}
\mathbf{q}_2 (\mathbf{k}, - \mathbf{q}) &=& \mathbf{k} - \frac{{\mathbf{q}}}
{2} + \frac{{\mathbf{k}} \cdot {\mathbf{q}}}
{{2\omega _k (2\omega _k  + {\bar M}_0 )}}\mathbf{q}
\label{a5_a}
\end{eqnarray}
and
\begin{eqnarray}
\mathbf{q}_3 (-\mathbf{k}, - \mathbf{q}) &=& -\mathbf{k} - \frac{{\mathbf{q}}}
{2} - \frac{{\mathbf{k} \cdot \mathbf{q}}}
{{2\omega _k (2\omega _k  + {\bar M}_0 )}}\mathbf{q} ~.
\label{a5_b}
\end{eqnarray}
That allows to write the three $\delta$-functions in the form
\begin{eqnarray}
&&  \delta (\mathbf{q} -  {\tilde \mathbf{q}} 
+ \frac{1}{3}(\mathbf{P} - \mathbf{P}') )
\delta (\mathbf{q}_2 (\mathbf{k}, - \mathbf{q}) - 
 {\tilde \mathbf{q}}_2^{~nr} + \frac{1}{3}(\mathbf{P} - \mathbf{P}') )
\delta(\mathbf{q}_3 ( - \mathbf{k}, - \mathbf{q}) 
-  {\tilde \mathbf{q}}_3^{~nr} + \frac{1}{3}(\mathbf{P} - \mathbf{P}') ) =   \cr
&&  \delta (\mathbf{q} - {\tilde \mathbf{q}} 
+ \frac{1}{3}(\mathbf{P} - \mathbf{P}') )
\delta ( -{\tilde \mathbf{p}} 
+ \mathbf{k} + \frac{{\mathbf{k} \cdot \mathbf{q}}}
{{2\omega _k (2\omega _k  + {\bar M}_0 )}}\mathbf{q} 
+ \frac{1}{3}(\mathbf{P} - \mathbf{P}') )\cr
&&\delta ( {\tilde \mathbf{p}} 
- \mathbf{k} - \frac{{\mathbf{k} \cdot \mathbf{q}}}
{{2\omega _k (2\omega _k  + {\bar M}_0 )}}\mathbf{q} 
+ \frac{1}{3}(\mathbf{P} - \mathbf{P}') ) = \cr
&&  \delta (\mathbf{q} - {\tilde \mathbf{q}} )
\delta ( {\tilde \mathbf{p}} 
- \mathbf{k} - \frac{{\mathbf{k} \cdot \mathbf{q}}}
{{2\omega _k (2\omega _k  + {\bar M}_0 )}}\mathbf{q})
\delta ( \mathbf{P} - \mathbf{P}' )  ~.
\label{a6}
\end{eqnarray}

In the following we assume the three-nucleon center of momentum system 
($\mathbf{P} = \mathbf{P}' = \mathbf{0}$) and drop the 
$\delta ( \mathbf{P} - \mathbf{P}' )$ in all expressions. 
 Performing the integration over ${d\hat {\tilde \mathbf{q}}}$ one gets
\begin{eqnarray}
  \langle k,q,\alpha  |
 \tilde p, \tilde q, \tilde \alpha  \rangle  &=&
 \frac{1} {2J+1} \sum\limits_{M}
 \sum\limits_{\mu_1 \mu_2 \mu_3 } \sum\limits_{\mu_2 '\mu_3 '} 
\sum\limits_{\mu_s \mu_l \mu_{\lambda}  \mu_I }
 \frac{{\delta (q - \tilde q)}}
{{q \tilde q}}
 \int d\hat \mathbf{q} \int d\hat \mathbf{k} 
  \int d\hat {\tilde \mathbf{p}} Y_{l\mu_l }^* (\hat \mathbf{k}) \cr
 && N(\mathbf{q}_2 ,\mathbf{q}_3 )
 (\frac {1}{2}, \frac {1}{2},s | \mu_2, \mu_3, \mu_s )
 (l,s,j | \mu_l, \mu_s, \mu ) \cr  
&&  D_{\mu_2 '\mu_2}^{1/2*} (R_{wc}(B_c ( -q),{k}_2( \mathbf{q}_2,\mathbf{q}_3 ))
D_{\mu_3 '\mu_3 }^{1/2*} 
(R_{wc}(B_c ( -q), {k}_3( \mathbf{q}_2,\mathbf{q}_3 ))  \cr
 && Y_{\lambda \mu_{\lambda}}^* (\hat \mathbf{q})(\lambda, \frac {1}{2},I |
 \mu_{\lambda}, \mu_1, \mu_I ) (j,I,J | \mu, \mu_I, M)  
  \sum\limits_{\tilde \nu_s, \tilde \nu_l, \tilde \nu_{\lambda}, \tilde \nu_I }
Y_{\tilde l \tilde \nu_l } 
(\hat {\tilde \mathbf{p}}) Y_{\tilde \lambda \tilde \nu_{\lambda}}
(\hat \mathbf{q})
 \cr
&&(\frac {1}{2}, \frac {1}{2}, \tilde s | \mu_2', \mu_3', \tilde {\nu}_s )
(\tilde l, \tilde s, \tilde j | \tilde \nu_l, \tilde \nu_s, \tilde \nu ) 
(\tilde \lambda, \frac {1}{2}, \tilde I | \tilde \nu_{\lambda}, \mu_1, 
\tilde \nu_I)
(\tilde j, \tilde I, J | \tilde \nu, \tilde \nu_I, M)   \cr
&&  \delta ( {\tilde \mathbf{p}} - \mathbf{k} - \frac{{\mathbf{k} 
\cdot \mathbf{q}}}
{{2\omega_k (2\omega_k  + {\bar M}_0 )}}\mathbf{q})  
  \langle (t \frac{1}{2})T| (\tilde t\frac{1}{2}) \tilde T \rangle  ~.
\label{a7}
\end{eqnarray}
That matrix element is a scalar which depends on the angles between
 the vectors 
$\mathbf{q}$, $\mathbf{k}$ and $ {\tilde \mathbf{p}}$. These angles are fixed by
the $\delta$-function $\delta ( {\tilde \mathbf{p}} - \mathbf{k} - 
\frac{{\mathbf{k} \cdot \mathbf{q}}}
{{2\omega _k (2\omega _k  + {\bar M}_0 )}}\mathbf{q})$. Namely, while
$ {\tilde \mathbf{p}} = \mathbf{k} + \frac{{\mathbf{k} \cdot \mathbf{q}}}
{{2\omega _k (2\omega _k  + {\bar M}_0 )}}\mathbf{q}$
it follows that
\begin{eqnarray}
\mathbf{k} \cdot {\tilde \mathbf{p}} &=& \mathbf{k} \cdot \mathbf{k} 
+ \frac{{(\mathbf{k} 
\cdot \mathbf{q})^2 }}
{{2\omega _k (2\omega _k  + {\bar M}_0 )}} ~,  \cr
\mathbf{q} \cdot {\tilde \mathbf{p}} &=& \mathbf{q} \cdot \mathbf{k} 
+ \frac{  ( {\mathbf{q} \cdot \mathbf{q}}) ( {\mathbf{k} \cdot \mathbf{q}}) }
{{2\omega _k (2\omega _k  + {\bar M}_0 )}} ~, \cr
{\tilde \mathbf{p}} \cdot {\tilde \mathbf{p}} &=& {\tilde \mathbf{p}} 
\cdot \mathbf{k} + 
\frac{{({\tilde \mathbf{p}} \cdot \mathbf{q})(\mathbf{k} \cdot \mathbf{q})}}
{{2\omega _k (2\omega _k  + {\bar M}_0 )}} ~.
\label{a7_a}
\end{eqnarray}
Therefore one can take $\hat \mathbf{q}$ pointing in z-direction, what for
given 
$k$, $q$, and $\tilde p$ values, defines all angles between the appearing
vectors. 
That allows to perform the integration over $d\hat \mathbf{q}$ resulting in
\begin{eqnarray}
  \langle k,q,\alpha  |
 \tilde p, \tilde q, \tilde \alpha  \rangle  &=&
 \frac{8\pi^2} {2J+1}
 \int\limits_{ - 1}^{ + 1} {dx}
 \frac{{\delta (q - \tilde q)}}
{{q\tilde q}}
\frac{{\delta (\tilde p - | {\mathbf{k} + \frac{{\mathbf{k} \cdot \mathbf{q}}}
{{2\omega_k (2\omega_k  + {\bar M}_0 )}}\mathbf{q}} |)}}
{{\tilde p^2 }} 
\sum\limits_{M}
 \sum\limits_{\mu_1 \mu_2 \mu_3 } \sum\limits_{\mu_2 '\mu_3 '} 
\sum\limits_{\mu_s \mu_l \mu_{\lambda}  \mu_I }
 Y_{l\mu_l }^* (\hat \mathbf{k}) \cr
 && N(\mathbf{q}_2 ,\mathbf{q}_3 )
 (\frac {1}{2}, \frac {1}{2},s | \mu_2, \mu_3, \mu_s ) 
(l,s,j | \mu_l, \mu_s, \mu )  \cr
&&  D_{\mu_2 '\mu_2 }^{1/2*} (R_{wc}(B_c ( -q),{k}_2( \mathbf{q}_2,\mathbf{q}_3 ))
D_{\mu_3 '\mu _3 }^{1/2*} (R_{wc}(B_c ( -q), {k}_3(
\mathbf{q}_2,\mathbf{q}_3 )) \cr
 && Y_{\lambda \mu _{\lambda}}^* (\hat \mathbf{q})(\lambda, \frac{1}{2}, I | 
\mu_{\lambda} , \mu_1, \mu_I ) (j,I,J | \mu, \mu_I, M) 
  \sum\limits_{\tilde \nu_s \tilde \nu_l \tilde \nu_{\lambda} \tilde \nu_I }
Y_{\tilde l\tilde \nu_l} 
(\hat {\tilde \mathbf{p}})Y_{\tilde \lambda \tilde \nu_{\lambda}}
(\hat \mathbf{q}) 
\cr
&&(\frac {1}{2}, \frac{1}{2}, \tilde s | \mu_2', \mu_3', \tilde {\nu}_s )
(\tilde l, \tilde s, \tilde j | \tilde \nu_l, \tilde \nu_s, \tilde \nu ) 
(\tilde \lambda, \frac {1}{2}, \tilde I | \tilde \nu_{\lambda},  \mu_1, 
\tilde \nu_I )
\cr 
&&(\tilde j, \tilde I,  J | \tilde \nu, \tilde \nu_I, M)  
  \langle (t, \frac{1}{2}),T| (\tilde t, \frac{1}{2}), \tilde T \rangle  
\label{a8}
\end{eqnarray}
with $x \equiv \hat \mathbf{q}\cdot \hat \mathbf{k}$.
We have chosen the coordinate system with $\mathbf{q}$ parallel to the
 z axis which leads to the 
  components of $\mathbf{q}$, $\mathbf{k}$, ${\tilde \mathbf{p}}$ 
\begin{eqnarray}
\mathbf{q} &=&(0, 0, q) ~, \cr
\mathbf{k} &=& (k\sqrt {1 - x^2 }, 0, kx) ~, \cr
{\tilde \mathbf{p}} &=& (k\sqrt {1 - x^2 } ,0,kx(1 + \frac{q^2 }
{2\omega _k (2\omega _k  + {\bar M}_0 )})) ~.
\label{a8_a}
\end{eqnarray}
The isospin factor is 
\begin{eqnarray}
\left\langle {{(t,\frac{1}
{2}),T}}  |
 {{(\tilde t,\frac{1}
{2}),\tilde T}} \right\rangle &=& 
\delta _{t\tilde t} \delta _{T\tilde T} \delta _{M_T M_{\tilde T} } 
\delta _{\nu _t \nu _{\tilde t} } ~.
\label{a8_b}
\end{eqnarray}
Taking that all together gives 
\begin{eqnarray}
\langle k,q,\alpha 
 | \tilde p, \tilde q, \tilde \alpha  \rangle  &=&
\delta_{t\tilde t} \delta_{T\tilde T}
\frac {2\pi \sqrt {(2\lambda  + 1)(2\tilde \lambda  + 1)}} {2J+1}
 \int\limits_{- 1}^{ + 1} dx \frac {{\delta (q - \tilde q)}} {{q\tilde q}}
\frac {\delta (\tilde p - | \mathbf{k} + \frac {\mathbf{k} \cdot \mathbf{q}}
{2\omega_k (2\omega_k  + {\bar M}_0 )}\mathbf{q} |)}
{\tilde p^2}  \cr
&&\sum\limits_{M}
 \sum\limits_{\mu_1 \mu_2 \mu_3} \sum\limits_{\mu_2 '\mu_3 '} 
 Y_{l,M-\mu_1-\mu_2-\mu_3}^* (\hat \mathbf{k}) N(\mathbf{q}_2 ,\mathbf{q}_3 ) \cr
 && D_{\mu_2 '\mu _2 }^{1/2*} (R_{wc}(B_c( -q),{k}_2( \mathbf{q}_2,\mathbf{q}_3 ))
D_{\mu_3 '\mu_3}^{1/2*} 
(R_{wc}(B_c( -q), -{k}_3( \mathbf{q}_2,\mathbf{q}_3 ))  \cr
 && (\frac {1}{2}, \frac {1}{2}, s | \mu_2, \mu_3, \mu_2 + \mu_3 )
(lsj | M-\mu_l -\mu_2-\mu_3, \mu_2+\mu_3, M-  \mu_1 ) \cr
 && (\lambda, \frac{1} {2},I | 0,  \mu_1, \mu_1 ) (j,I,J | M-\mu_1, \mu_1, M)
Y_{\tilde l,M- \mu_1 - \mu_2 '-\mu_3 '} (\hat {\tilde \mathbf{p}}) \cr
&&(\frac {1} {2}, \frac {1} {2}, \tilde s | \mu_2 ', \mu_3 ', \mu_2
'+\mu_3 ')
(\tilde l, \tilde s, \tilde j | M-\mu_1-\mu_2 '-\mu_3 ', \mu_2 '+\mu_3 ',
  M-\mu_1)  \cr
&& (\tilde \lambda, \frac {1}{2}, \tilde I | 0, \mu_1 \mu_1) 
(\tilde j, \tilde I,  J | M-\mu_1, \mu_1, M )    ~.
\label{a9}
\end{eqnarray}
The resulting expression for the matrix element 
$\left\langle {k,q,\alpha } \right|V_4^{(1)} (1 + P)\left| {k,'q',\alpha
  '}\right\rangle$ is given by
\begin{eqnarray}
&&\langle k,q,\alpha  | V_4^{(1)} (1 + P)
 | k',q',\alpha ' \rangle 
  =  
\sum\limits_{\tilde {\alpha}}  
  \delta_{t \tilde t} \delta_{T\tilde T} \frac {2\pi 
\sqrt {(2\lambda  + 1)(2\tilde \lambda  + 1)}} {2J+1}  \cr 
&& \int\limits_{ - 1}^{ + 1} dx \sum\limits_{M}
 \sum\limits_{\mu_1 \mu_2 \mu_3 } \sum\limits_{\bar {\mu}_2 \bar {\mu}_3 }
 Y_{l,M-\mu_1-\mu_2-\mu_3}^* (\hat \mathbf{k}) N(\mathbf{q}_2 ,\mathbf{q}_3 ) \cr
&&  D_{\bar {\mu_2} \mu_2}^{1/2*} (R_{wc}(B_c ( -q),
{k}_2( \mathbf{q}_2,\mathbf{q}_3 ))
D_{\bar {\mu_3} \mu_3}^{1/2*} 
(R_{wc}(B_c ( -q), {k}_3( \mathbf{q}_2,\mathbf{q}_3 ))  \cr
 && (\frac {1}{2}, \frac{1}{2}, s | \mu_2, \mu_3, \mu_2 + \mu_3 )
(l,s,j | M-\mu_l -\mu_2-\mu_3, \mu_2+\mu_3, M-  \mu_1 )  \cr
 && (\lambda, \frac {1}{2}, I | 0,  \mu_1, \mu_1 ) (j,I,J | M-\mu_1, \mu_1, M)
Y_{\tilde l,M- \mu_1 - {\bar \mu}_2 - {\bar \mu}_3} (\hat {\tilde \mathbf{p}}) \cr
&&(\frac {1}{2}, \frac {1}{2}, \tilde s | {\bar \mu}_2, {\bar \mu}_3, 
{\bar \mu}_2+{\bar \mu}_3 ) (\tilde l, \tilde s, \tilde j | 
M-\mu_1- {\bar \mu}_2-{\bar \mu}_3, {\bar \mu}_2+{\bar \mu}_3, M-\mu_1)  \cr
&& (\tilde \lambda, \frac {1}{2}, \tilde I | 0, \mu_1, \mu_1) 
(\tilde j, \tilde I,  J | M-\mu_1, \mu_1, M)   
 \sum\limits_{\tilde {\alpha}'}
  \delta _{t'\tilde t'} \delta _{T'\tilde T'} \frac {2\pi 
\sqrt {(2\lambda'  + 1)(2\tilde \lambda'  + 1)}} {2J+1}  \cr
&& \int\limits_{ - 1}^{ + 1} dx'
 \sum\limits_{M}
 \sum\limits_{\mu_1' \mu_2' \mu_3' } \sum\limits_{ {\bar \mu}_2' {\bar \mu}_3' }
 Y_{l',M-\mu_1'-\mu_2'-\mu_3'} (\hat \mathbf{k}') N(\mathbf{q}_2'
 ,\mathbf{q}_3' )
\cr && 
  D_{{\bar \mu}_2' \mu_2'}^{1/2} (R_{wc}(B_c( -q'),
{k}_2'( \mathbf{q}_2',\mathbf{q}_3' ))
D_{{\bar \mu}_3' \mu _3'}^{1/2} (R_{wc}(B_c( -q'), 
{k}_3'( \mathbf{q}_2',\mathbf{q}_3' ))  \cr
 && (\frac {1}{2}, \frac {1}{2},s' | \mu_2', \mu_3', \mu_2' +
  \mu_3') (l',s',j' | M-\mu_l' -\mu_2'-\mu_3', \mu_2'+\mu_3', M-  \mu_1' )  \cr
 && (\lambda', \frac {1}{2},I' 
| 0 , \mu_1', \mu_1' ) (j',I',J | M-\mu_1', \mu_1', M)  
Y_{\tilde l',M- \mu_1'-{\bar \mu}_2' - {\bar \mu}_3'} (\hat {\tilde \mathbf{p}}') \cr
&&(\frac {1}{2}, \frac{1}{2}, \tilde s' | {\bar \mu}_2', {\bar \mu}_3', 
{\bar \mu}_2'+{\bar \mu}_3' ) 
(\tilde l', \tilde s', \tilde j' | M-\mu_1'-{\bar \mu}_2'-{\bar \mu}_3', 
{\bar \mu}_2'+{\bar \mu}_3', M-\mu_1'  )   \cr
&& (\tilde \lambda', \frac {1}{2}, \tilde I' | 0, \mu_1', \mu_1') 
(\tilde j', \tilde I',  J | M-\mu_1', \mu_1', M )    \cr
&& \langle | \mathbf{k} + \frac {\mathbf{k} \cdot \mathbf{q}}
{2\omega_k (2\omega_k  + {\bar M}_0 )} \mathbf{q}~ |,
 q, \alpha   |V_4^{(1)} (1 + P) | | \mathbf{k}' + 
\frac{\mathbf{k}' \cdot \mathbf{q}'}
 {2\omega_{k'} (2\omega_{k'}  + {\bar M}_0 )} \mathbf{q}' |
, q', \tilde {\alpha}' \rangle ~.
\label{a10}
\end{eqnarray}

\section{Transformation of a 3-dimensional three-nucleon 
force matrix element from
  ($\mathbf{p}, \mathbf{q}$) to ($\mathbf{k}, \mathbf{q}$) momenta}
\label{app_b}

We would like to express directly the matrix element
$\left\langle {\mathbf{k}, \mathbf{q}}~ \right|V_4^{(1)} 
\left| {\mathbf{k}'}, \mathbf{q}' \right\rangle$ by that 
matrix element given in terms of single-nucleon momenta
$\left\langle {\mathbf{q}_1, \mathbf{q}_2, \mathbf{q}_3} 
\right|V_4^{(1)} \left| {\mathbf{q}_1', \mathbf{q}_2', \mathbf{q}_3'} 
\right\rangle$.
 For momenta $\mathbf{q}_2$ and $\mathbf{q}_3$ of 
nucleons 2 and 3 in three-nucleon center of momentum system their 
relative momentum
 $\mathbf{k}(\mathbf{q}_2 ,\mathbf{q}_3 )$ in the two-nucleon center 
of momentum subsystem 
of nucleons 2 and 3 is
\begin{eqnarray}
\mathbf{k}(\mathbf{q}_2 ,\mathbf{q}_3 ) = \frac{1}
{2}[\mathbf{q}_2  - \mathbf{q}_3  - (\mathbf{q}_2  + \mathbf{q}_3 )
\frac{{E_2  - E_3 }}
{{E_2  + E_3  + \sqrt {(E_2  + E_3 )^2  - (\mathbf{q}_2  
+ \mathbf{q}_3 )^2 } }}]
\label{equ.B1}
\end{eqnarray}
with $E_i  = \sqrt {m^2  + q_i^{~2} }$.

Using completeness of $\left| {\mathbf{q}_1 \mathbf{q}_2 \mathbf{q}_3}
\right\rangle$ 
states one gets
\begin{eqnarray}
  \langle \mathbf{k},\mathbf{q}  | V_4^{(1)} | \mathbf{k}',\mathbf{q}' 
\rangle  &=& \int d\mathbf{q}_1  d\mathbf{q}_2 d\mathbf{q}_3 
\delta (\mathbf{q}_1  + \mathbf{q}_2  + \mathbf{q}_3 ) \langle
\mathbf{k}, \mathbf{q}  | 
 \mathbf{q}_1, \mathbf{q}_2, \mathbf{q}_3 \rangle 
\langle \mathbf{q}_1, \mathbf{q}_2, \mathbf{q}_3 |  V_4^{(1)} \cr
&& \int d\mathbf{q}_1' d\mathbf{q}_2'd\mathbf{q}_3' | \mathbf{q}_1'
\mathbf{q}_2'\mathbf{q}_3' \rangle \langle \mathbf{q}_1',
\mathbf{q}_2', \mathbf{q}_3'| 
\mathbf{k}', \mathbf{q}'\rangle \delta (\mathbf{q}_1' + \mathbf{q}_2' + 
\mathbf{q}_3')  \cr
&=&  \int d\mathbf{q}_1  d\mathbf{q}_2 d\mathbf{q}_3 \delta
(\mathbf{q}_1  
+ \mathbf{q}_2  
+ \mathbf{q}_3 ) \delta (\mathbf{q} - \mathbf{q}_1 ) 
\delta (\mathbf{k} - \mathbf{k}(\mathbf{q}_2 ,
\mathbf{q}_3 ))  \cr
&&  \int d\mathbf{q}_1' d\mathbf{q}_2'd\mathbf{q}_3'
\delta (\mathbf{q}_1' + \mathbf{q}_2' 
+ \mathbf{q}_3')\delta (\mathbf{q}' - \mathbf{q}_1')\delta (\mathbf{k}' 
- \mathbf{k}'(\mathbf{q}_2',\mathbf{q}_3'))  \cr
&&  \frac{1}{N(\mathbf{q}_2 ,\mathbf{q}_3 )} \frac{1}
{N(\mathbf{q}_2',\mathbf{q}_3')} \langle \mathbf{q}_1 \mathbf{q}_2 
\mathbf{q}_3 |V_4^{(1)} | \mathbf{q}_1'\mathbf{q}_2'\mathbf{q}_3' \rangle   \cr
&&   = \int d\mathbf{q}_2 d\mathbf{q}_3 \delta (\mathbf{q} +
\mathbf{q}_2  + \mathbf{q}_3 )
\delta (\mathbf{k} - \mathbf{k}(\mathbf{q}_2 ,\mathbf{q}_3 ))   \cr
&&  \int d\mathbf{q}_2'd\mathbf{q}_3'\delta (\mathbf{q}' +
\mathbf{q}_2' + \mathbf{q}_3')\delta (\mathbf{k}' -
\mathbf{k}'(\mathbf{q}_2'
,\mathbf{q}_3'))   \cr
&&  \frac{1}{N(\mathbf{q}_2 ,\mathbf{q}_3 )} \frac{1}
{N(\mathbf{q}_2',\mathbf{q}_3')} \langle \mathbf{q},\mathbf{q}_2 ,\mathbf{q}_3 
 |V_4^{(1)} | \mathbf{q}',\mathbf{q}_2',\mathbf{q}_3' \rangle   \cr
&&   = \int d\mathbf{k}(\mathbf{q}_2 ,\mathbf{q}_3 )
 d(\mathbf{q}_2+\mathbf{q}_3) \delta (\mathbf{q} +
\mathbf{q}_2  + \mathbf{q}_3 )
\delta (\mathbf{k} - \mathbf{k}(\mathbf{q}_2 ,\mathbf{q}_3 ))   \cr
&&  \int d\mathbf{k}'(\mathbf{q}_2',\mathbf{q}_3')
d(\mathbf{q}_2'+\mathbf{q}_3') \delta (\mathbf{q}' +
\mathbf{q}_2' + \mathbf{q}_3')\delta (\mathbf{k}' -
\mathbf{k}'(\mathbf{q}_2'
,\mathbf{q}_3'))   \cr
&&  {N(\mathbf{q}_2 ,\mathbf{q}_3 )} 
{N(\mathbf{q}_2',\mathbf{q}_3')} \langle \mathbf{q},\mathbf{q}_2 ,\mathbf{q}_3 
 |V_4^{(1)} | \mathbf{q}',\mathbf{q}_2',\mathbf{q}_3' \rangle   ~.
\label{transf_0}
\end{eqnarray}
For given vectors $\mathbf{k}(\mathbf{q}_2,\mathbf{q}_3)=\mathbf{k}_0$ 
 and $\mathbf{q}_2+\mathbf{q}_3=-\mathbf{q}_0$ 
 the vectors $\mathbf{q}_2$ and $\mathbf{q}_3 $  are given by: 
$\mathbf{q}_2=-\mathbf{q}_0-\mathbf{q}_3^0$ and
$\mathbf{q}_3=\mathbf{q}_3^0$ 
 with $\mathbf{q}_3^{0}$ being the solution of
 the equation
\begin{eqnarray}
\mathbf{k}_0 - \mathbf{k}( - \mathbf{q}_0 - \mathbf{q}_3^{0}
,\mathbf{q}_3^{0} ) = \mathbf{ 0}
\label{solve_p30}
\end{eqnarray}
and similarily for the primed quantities.

Thus one gets
\begin{eqnarray}
  \left\langle {\mathbf{k}\mathbf{q}} \right|V_4^{(1)} 
\left| {\mathbf{k}'\mathbf{q}'} \right\rangle  &=& 
{{N( - \mathbf{q} - \mathbf{q}_3^{0} ,\mathbf{q}_3^{0} )}} 
{{N( - \mathbf{q}' - {\mathbf{q}_3^0}',{\mathbf{q}_3^0}')}}  \cr
&&  \left\langle {\mathbf{q}, - \mathbf{q} - \mathbf{q}_3^{0} ,
\mathbf{q}_3^{0} } \right|V_4^{(1)} \left| {\mathbf{q}', - \mathbf{q}' 
- {\mathbf{q}_3^0}',{\mathbf{q}_3^0}'} \right\rangle   \cr
&=& {{N( - \mathbf{q} - \mathbf{q}_3^{0} ,\mathbf{q}_3^{0} )}} 
{{N( - \mathbf{q}' - {\mathbf{q}_3^0}',{\mathbf{q}_3^0}')}}  \cr
&&  \left\langle {\mathbf{p} = -\frac {1} {2}\mathbf{q} - \mathbf{q}_3^{0} ,
\mathbf{q} } \right|V_4^{(1)} \left| {\mathbf{p}'= - \frac {1} {2}\mathbf{q}' 
- {\mathbf{q}_3^0}',\mathbf{q}'} \right\rangle ~.
\label{transf_1}
\end{eqnarray}

Starting from (\ref{b.10}) and following the same steps as in
(\ref{transf_0}) one gets for the partial wave projected matrix elements
\begin{eqnarray}
  \langle k,q,\alpha  | V_4^{(1)} |k',q',\alpha' \rangle  &=&
\int d\hat{\mathbf{k}}  d\hat{\mathbf{q}}_1    
  \sum_{ \bar{\mu}_2 \bar{\mu}_3 } \sum_{\mu_2 \mu_3 \mu_s }
 \sum_{\mu_l \mu_{\lambda} \mu_I }
  ( {1\over{2}}, \mu_2 ,{1\over{2}}, \mu_3 \vert s, \mu_s )
 (  l,  \mu_l, s, \mu_s, \vert j, \mu_j ) \cr
&~& (\lambda, \mu_{\lambda}, {1\over{2}}, \mu_1 \vert I, \mu_I )
( j, \mu_j, I, \mu_I \vert J, \mu) \
 Y^*_{\lambda \mu_{\lambda}}(\hat{\mathbf{q}}_1)
 Y^*_{l \mu_l}(\hat{\mathbf{k}} )  \cr
&~&
D^{{1\over{2}} * }_{\bar{\mu}_2 \mu_2}
[R_{wc}(~ B_c(-q_1),
{k}_2(~ \mathbf{q}_2,\mathbf{q}_3 )~ )] \cr 
&~& D^{{1\over{2}} * }_{\bar{\mu}_3 \mu_3}
[R_{wc}(~ B_c(-q_1),
{k}_3(~ \mathbf{q}_2,\mathbf{q}_3 )~ )] \cr
&&  \int d\hat{\mathbf{k}}' d\hat{\mathbf{q}}_1'    
 \sum_{ \bar{\mu}_2' \bar{\mu}_3' } \sum_{\mu_2' \mu_3' \mu_s' }
 \sum_{\mu_l' \mu_{\lambda}' \mu_I' }
  ( {1\over{2}}, \mu_2' ,{1\over{2}}, \mu_3' \vert s', \mu_s' )
 (  l',  \mu_l', s', \mu_s', \vert j', \mu_j' ) \cr
&~& (\lambda', \mu_{\lambda}', {1\over{2}}, \mu_1' \vert I', \mu_I' )
( j', \mu_j', I', \mu_I' \vert J, \mu) 
 Y_{\lambda' \mu_{\lambda}'}(\hat{\mathbf{q}}_1')
 Y_{l' \mu_l'}( \hat{\mathbf{k}}' )  \cr
&~&
D^{1\over{2} }_{\bar{\mu}'_2 \mu_2'}
[R_{wc}(~ B_c(-q_1'),
{k}_2'(~ \mathbf{q}_2',\mathbf{q}_3' )~ )]  \cr
&~& D^{1\over{2} }_{\bar{\mu}_3' \mu_3'}
[R_{wc}(~ B_c(-q_1'),
{k}_3(~ \mathbf{q}_2',\mathbf{q}_3' )~ )] \cr
&&  {N(\mathbf{q}_2 ,\mathbf{q}_3 )} 
{N(\mathbf{q}_2',\mathbf{q}_3')} \langle \mathbf{q},\mathbf{q}_2 ,\mathbf{q}_3 
 |V_4^{(1)} | \mathbf{q}',\mathbf{q}_2',\mathbf{q}_3' \rangle   ~,
\label{transf_2}
\end{eqnarray}
where $\mathbf{q}_1 \equiv q \hat{\mathbf{q}}_1$,   
$\mathbf{k}_2(\mathbf{q}_2,\mathbf{q}_3) \equiv k 
\hat{\mathbf{k}}$, 
 $\mathbf{k}_3(\mathbf{q}_2,\mathbf{q}_3)= 
-\mathbf{k}_2(\mathbf{q}_2,\mathbf{q}_3)$, and $\mathbf{q}_2$
together with $\mathbf{q}_3$ result from (\ref{solve_p30}) and
similarily for primed quantities. These partial wave
matrix elements can be obtained using automatized partial wave
expansion of Ref. \cite{automa2}.

\newpage

\newpage 

\end{document}